\documentclass[english,12pt]{article}
\usepackage[T1]{fontenc}
\usepackage[utf8]{inputenc}
\usepackage[pdftex]{graphicx}
\usepackage[english]{babel}
\usepackage{verbatim}
\usepackage{lmodern}
\usepackage[width=\textwidth]{caption}
\usepackage{amsmath,amssymb,amsopn,dsfont,bm,mathrsfs}
\usepackage{a4wide}
\usepackage{float, placeins,flafter,lscape,longtable,array,booktabs}
\usepackage{color}
\usepackage{natbib}
\usepackage{zref-base}
\usepackage{atveryend}
\makeatletter
\AfterLastShipout{%
  \if@filesw
    \begingroup
      \zref@setcurrent{default}{\the\value{equation}}%
      \zref@wrapper@immediate{%
        \zref@labelbyprops{eq-last}{default}%
      }%
      \zref@setcurrent{default}{\the\value{hyp}}%
      \zref@wrapper@immediate{%
        \zref@labelbyprops{hyp-last}{default}%
      }%
      \zref@setcurrent{default}{\the\value{thm}}%
      \zref@wrapper@immediate{%
        \zref@labelbyprops{thm-last}{default}%
      }%
    \endgroup
  \fi
}
\makeatother

\makeatletter

\newtheorem{thm}{Theorem}
\newtheorem{cor}{Corollary}
\newtheorem{prop}{Proposition}
\newtheorem{lem}{Lemma}
\newtheorem{hyp}{Assumption}
\newtheorem{mydef}{Regression}

\newcommand{\R}{\mathbb R}

\newcommand{\indic}[1]{\mathds{1}\left\{#1\right\}}

\newcommand{\eps}{\varepsilon}

\newcommand{\convP}{\stackrel{P}{\longrightarrow}}
\newcommand{\convL}{\stackrel{d}{\longrightarrow}}
\newcommand{\sgn}{\text{sgn}}
\newcommand{\st}[1]{\texttt{#1}}

\newcommand{\DID}{\text{DID}}
\newcommand{\DIDM}{\DID_{\text{M}}}

\newcommand{\0}{\bm{0}}
\newcommand{\1}{\bm{1}}
\newcommand{\CIex}{\text{CI}^{\,\text{ex}}_{1-\alpha}}
\newcommand{\CI}{\text{CI}_{1-\alpha}}

\renewcommand{\section}{\@startsection{section}{2}{0mm}{-1.5\baselineskip}{1\baselineskip}{\normalfont\large\bfseries}}
\renewcommand{\subsection}{\@startsection{subsection}{2}{0mm}{-1.2\baselineskip}{1\baselineskip}{\normalfont\normalsize\bfseries}}
\renewcommand{\subsubsection}{\@startsection{subsubsection}{3}{0mm}{-0.8\baselineskip}{0.4\baselineskip}{\normalfont\normalsize\itshape}}

\date{}
\begin{document}

\title{Two-way Fixed Effects and Differences-in-Differences Estimators with Several Treatments\thanks{Several of this paper's ideas arose during conversations with Enrico Cantoni, Angelica Meinhofer, Vincent Pons, Jimena Rico-Straffon, Marc Sangnier, Oliver Vanden Eynde, and Liam Wren-Lewis who shared with us their interrogations, and sometimes their referees' interrogations, on two-way fixed effects regressions with several treatments. We are grateful to them for those stimulating conversations. We are grateful to Yubo Wei for his excellent work as a research assistant. We are also grateful to the editor, associate editor, and two anonymous referees for their very helpful comments.}}

\author{Cl\'{e}ment de Chaisemartin\and Xavier D'Haultf\oe{}uille\thanks{de Chaisemartin: Sciences Po (email: clement.dechaisemartin@sciencespo.fr); D'Haultf\oe uille: CREST-ENSAE (email: xavier.dhaultfoeuille@ensae.fr). Xavier D'Haultf\oe uille thanks the hospitality of PSE where this research was conducted.}}

\maketitle ~\vspace{-1cm}

\begin{abstract}
We study two-way-fixed-effects regressions (TWFE) with several treatment variables. Under a parallel trends assumption, we show that the coefficient on each treatment identifies a weighted sum of that treatment’s effect, with possibly negative weights, plus a weighted sum of the effects of the other treatments. Thus, those estimators are not robust to heterogeneous effects and may be contaminated by other treatments’ effects. We further show that omitting a treatment from the regression can actually reduce the estimator's bias, unlike what would happen under constant treatment effects. We propose an alternative difference-in-differences estimator, robust to heterogeneous effects and immune to the contamination problem. In the application we consider, the TWFE regression identifies a highly non-convex combination of effects, with large contamination weights, and one of its coefficients significantly differs from our heterogeneity-robust estimator.

\textit{(JEL C21, C23)}
\end{abstract}

\section{Introduction}

To estimate treatment effects, researchers often use panels of groups (e.g. counties, regions), and estimate two-way fixed effect (TWFE) regressions, namely regressions of the outcome variable on group and time fixed effects and the treatment. \cite{dechaisemartin2020two} have found that almost 20\% of empirical papers published by the American Economic Review (AER) from 2010 to 2012 estimate such regressions.

\medskip
Under a parallel trends assumption, TWFE regressions with one treatment identify a weighted sum of the treatment effects of treated $(g,t)$ cells, with weights that may be negative and sum to one \citep[see][]{dechaisemartin2020two,borusyak2016}. Because of the negative weights, the treatment coefficient in such regressions is not robust to heterogeneous treatment effects across groups and time periods: it may be, say, negative, even if the treatment effect is strictly positive in every $(g,t)$ cell.

\medskip
However, in 18\% of the TWFE papers published in the AER from 2010 to 2012, the TWFE regression has several treatment variables. By including several treatments, researchers hope to estimate the effect of each treatment holding the other treatments constant. For instance, when studying the effect of marijuana laws, as in \cite{meinhofer2021marijuana}, one may want to separate the effect of medical and recreational laws. To do so, one may estimate a regression of the outcome of interest in state $g$ and year $t$ on state fixed effects, year fixed effects, an indicator for whether state $g$ has a medical law in year $t$, and an indicator for whether state $g$ has a recreational law in year $t$.

\medskip
In this paper, we investigate what TWFE regressions with several treatments identify. We show that under a parallel trends assumption, the coefficient on each treatment identifies the sum of two terms. The first term is a weighted sum of the effect of that treatment in each group and period, with weights that may be negative and sum to one. A similar weighted sum appears in decompositions of TWFE regressions with only one treatment.
The second term is a sum of the effects of the other treatments, with weights summing to zero. Accordingly, with several treatments, coefficients in TWFE regressions may be contaminated by the effect of other treatments, an issue that was not present with one treatment. As the weights sum to zero, this second term disappears if the effect of the other treatments is homogeneous, but it is often implausible that those effects are homogeneous. The weights attached to any TWFE regression with several treatments can be computed by the \st{twowayfeweights} Stata and R packages. Estimating those weights may be useful, to assess if a TWFE coefficient is robust to heterogeneous treatment effects, and if it is contaminated by the effect of the other treatments in the regression.

\medskip
We consider simple examples with two treatments, to show that TWFE regressions may not be robust to heterogeneous effects because they may leverage two types of ``forbidden comparisons'', borrowing the terminology coined by \cite{borusyak2016}. In a first example, the coefficient on the first treatment leverages a difference-in-differences (DID) comparing the outcome evolution of a group going from untreated to receiving both treatments to the outcome evolution of a ``control'' group going from untreated to receiving the second treatment. If the effect of the second treatment is the same in the two groups, those two effects cancel each other out in this DID. But if the effects of the second treatment differ in the two groups, they do not cancel each other out, and they contaminate the coefficient on the first treatment. In a second example, the coefficient on the first treatment leverages a DID comparing the outcome evolution of a group going from untreated to receiving the first treatment to the outcome evolution of a ``control'' group that receives the second treatment at both periods. If the control group's effect of the second treatment is the same in the pre and in the post period, those two effects cancel each other out in this DID. But if the control group's effect of the second treatment changes over time, those two effects do not cancel out, and they contaminate the coefficient on the first treatment.

\medskip
We then consider a TWFE regression that would omit the other treatments, and derive another decomposition formula. In the presence of two treatments, a TWFE regression with only the first treatment also estimates a weighted sum of the effect of that treatment in each group and period, with weights that may be negative and sum to one, plus a weighted sum of the effects of the other treatments, but with weights that do not sum to zero. Then, we use our decompositions of the TWFE regressions with one and several treatments to derive the maximal bias of both regressions for the average effect of the first treatment on the treated, under the assumption that the effect of every treatment is bounded in absolute value by a (potentially large) constant in every group and period. The ratio between the maximal biases of both regressions is independent of that constant and can be estimated, thus allowing researchers to compare the maximal bias of the two regressions. The ratio of the regressions' maximal biases can either be smaller or larger than one in practice, which means that controlling for more treatments may lead to a more biased estimator than not controlling for them. Therefore, omitting a treatment from the regression can actually reduce the estimator's bias, something that cannot happen under constant treatment effects.

\medskip
Finally, we propose an alternative DID estimator that relies on common trends assumptions, like TWFE regressions, but that is robust to heterogeneous effects and does not suffer from the contamination problem, unlike TWFE regressions. Our estimator generalizes the $\DIDM$ estimator in \cite{dechaisemartin2020two} to instances with several treatments. To isolate the effect of the first treatment, our estimator compares the $t-1$-to-$t$ outcome evolution, of switching groups whose first treatment switches from $t-1$ to $t$ while their other treatments do not change, and of control groups i) whose treatments all remain the same, and ii) that had the same treatments as the switching groups in period $t-1$. i) ensures that our new estimator is robust to heterogeneous effects across groups of all treatments. ii) ensures that it is robust to heterogeneous effects over time of all treatments.

\medskip
Our estimator's robustness may come at a high price in terms of external validity and statistical precision. For instance, in our application in Section \ref{sec:appli}, we can only match a small number of switchers to valid control groups meeting i) and ii). Then, there may be internal-external validity and bias-variance trade-offs between our new estimator and less robust estimators, such as the $\DIDM$ estimator in \cite{dechaisemartin2020two} or TWFE regressions with several treatments. To account for the fact our new estimator may sometimes be estimated on a small sample of groups, we propose,
in addition to a standard confidence interval  that is asymptotically valid under weak conditions, another confidence interval that has both exact coverage under a normality assumption and is asymptotically valid without such a normality requirement.

\medskip
As an illustration, we use our results to revisit \cite{hotz2011impact}, who run TWFE regressions of measures of daycare quality in state $g$ and year $t$ on two daycare regulations in state $g$ and year $t$: the minimum number of years of schooling required to be a daycare director and the minimum staff-to-child ratio. Focusing on the years-of-schooling treatment, we find that the TWFE regression with several treatments estimates weighted sums of effects with very large negative weights attached to them, both on the treatment's own effects, but also on the effects of the other treatments in the regression. The TWFE regression with only the years-of-schooling treatment has much smaller weights attached to it.
As a result, the maximal bias of the TWFE regression with several treatments is almost five times larger than that of the regression including only the years-of-schooling treatment. Thus, the ``short'' regression seems preferable, at least per our maximal-bias metric. We finally show that our heterogeneity-robust estimator is much closer to zero than, and significantly different from, the coefficient of the TWFE regression with several treatments.

\medskip
The remainder of the paper is organized as follows. Section 2 discusses the related literature. Section 3 presents the set up. Section 4 presents our decomposition results for TWFE regressions with several treatments. Section 5 presents our alternative estimator. Section 6 presents our empirical application.

\section{Related literature} 
\label{sub:literature}

\subsection{Connection with papers studying TWFE regressions with one treatment}

Our paper is closely related to the recent literature showing that TWFE regressions with one treatment variable may not be robust to heterogeneous effects \citep[see][]{dechaisemartin2020two,goodman2021difference,borusyak2016}. In particular, Theorem S4 in the Web Appendix of \cite{dechaisemartin2020two} studies TWFE regressions with one treatment and some time-varying control variables. Under a parallel trend assumption accounting for such covariates, \cite{dechaisemartin2020two} show that TWFE regressions with one treatment and some controls identify a weighted sum of the treatment effects across all treated $(g,t)$ cells. Our decomposition results are related to, but different from, that result. The weighted sum in Theorem S4 of \cite{dechaisemartin2020two} is identical to the first weighted sum in Theorem \ref{thm:extension} below. On the other hand, the second weighted sum in Theorem \ref{thm:extension}, the contamination term, does not appear in Theorem S4 of \cite{dechaisemartin2020two}. This is because the parallel trend assumptions are not the same in the two results. When the other variables in the regression are treatments rather than covariates (see below for the difference between a treatment and a covariate), one can show that the parallel trends condition in Theorem S4 implicitly assumes that the effect of the other treatments is constant, which is why the contamination term disappears.

\subsection{Connection with papers studying linear regressions with several treatments}

Our paper is also related to three papers that also consider linear regressions with several treatments.

\medskip
Firstly, our results complement the pioneering work of \cite{abraham2018}. The authors study the so-called event-study regression, an example of a TWFE regression with several treatments, where the treatments are indicators for having started receiving a single binary-and-staggered treatment $\ell$ periods ago. In those regressions, the authors show that effects of being treated for $\ell'$ periods may contaminate the coefficient supposed to measure the effect of $\ell$ periods of treatment in the regression, and they provide a decomposition formula one can use to quantify the extent of the phenomenon. If i) the $K$ treatments we consider are indicators for having started receiving a single binary-and-staggered treatment $\ell$ periods ago, and ii) the treatment no longer has an effect after $K+1$ periods of exposure, then our Theorem \ref{thm:extension} reduces to Proposition 3 in \cite{abraham2018}, provided no lags are gathered together in the event-study regression they consider.\footnote{In their decomposition, \cite{abraham2018} gather groups that started receiving the treatment at the same period into cohorts. Their decomposition can then be further decomposed, 
finally leading to the result in our Theorem \ref{thm:extension}.}

\medskip
Our decompositions extend their result, by showing that the contamination bias they first uncovered is very pervasive: it can arise in any TWFE regression with several treatments, rather than in event-study regressions only. In particular, our results apply to situations where the treatments are different, potentially non-mutually exclusive policies, that may not be binary or may not follow a staggered adoption design. Another difference with their work is that with non-mutually exclusive treatments, the contamination weights do not sum to zero. We also provide some novel intuition as to why contamination may arise with different, potentially non-mutually exclusive treatments in the regression. Finally, we show that omitting the other treatments from the regression may not necessarily increase the regression coefficient's bias.

\medskip
Secondly, Theorem 1 is also related to the pioneering work of \cite{hull2018}. In his Section 2.2, the author studies TWFE regressions where indicators for each value that a multinomial treatment may take are included in the regression, an example of a TWFE regression with several treatments. Equation (15) therein is, to our knowledge, the first instance where a contamination phenomenon was shown. However, the paper does not discuss this phenomenon. It also does not give a decomposition formula, so one cannot use the paper's results to compute the contamination weights, and assess whether they are important in a given regression. Finally, the paper's result applies when the data has two periods, and in instances were the treatments in the regression are indicators for each value that a multinomial treatment may take.

\medskip
Thirdly, another related paper, released after ours, is \cite{goldsmith2021estimating}, who show that a contamination phenomenon similar to that in \cite{abraham2018} and in Theorems \ref{thm:main} and \ref{thm:extension} below also arises in linear regressions with several treatments, and a set of controls such that the treatments can be assumed to be independent of the potential outcomes conditional on those controls. Their result is not nested within and does not nest the results of \cite{abraham2018} nor ours: both \cite{abraham2018} and us assume parallel trends rather than conditional independence. The weights in their decomposition are functions of the variance-covariance matrix of the treatments conditional on the controls. An interesting difference with our results is that under their conditional independence assumption, the weights on the effect of the first treatment are all positive.

\medskip
Overall, our four papers complement each other, and show that the contamination phenomenon is very pervasive, as it arises under several identifying assumptions (parallel trends and conditional independence), and irrespective of the nature of the treatments included in the regression.

\section{Set up}

We consider a panel of $G$ groups observed at $T$ periods, respectively indexed by $g$ and $t$. Typically, groups are geographical entities gathering many observations, but a group could also just be a single individual or firm. For every $(g,t)\in \{1,...,G\}\times \{1,...,T\}$, let $N_{g,t}$ denote the population of cell $(g,t)$, and let $N=\sum_{g,t}N_{g,t}$ be the total population across all cells.

\medskip
We are interested in the effect of $K$ treatments. In this paper, we follow, e.g., \cite{holland1986statistics,holland1987causal}, and define as a treatment a variable that has a causal effect on the outcome, in the sense that different values of that variable lead to different counterfactual outcomes.\footnote{By contrast, a covariate may be statistically correlated to the outcome but does not have a causal effect on it.}
For every $(k,g,t)\in \{1,...,K\}\times \{1,...,G\}\times \{1,...,T\}$, let $D^k_{g,t}$ denote the value of treatment $k$ for group $g$ at period $t$, and let $\bm{D}_{g,t}=(D^k_{g,t})_{k\in \{1,...,K\}}$ denote a vector stacking together the $K$ treatments of group $g$ at period $t$. For every $k$, let $\mathcal{D}_k$ denote the values $D^k_{g,t}$ can take. For now, we assume that the treatments are binary: $\mathcal{D}_k=\{0,1\}$ for all $k$. This is just to simplify the exposition: our results can be extended to non-binary treatments, as explained below.

\medskip
For any $\bm{d}\in \{0,1\}^K$, let $Y_{g,t}(\bm{d})$ denote the potential outcome of group $g$ at period $t$ if $\bm{D}_{g,t}=\bm{d}$. The observed outcome is $Y_{g,t}=Y_{g,t}(\bm{D}_{g,t})$. Importantly, our notation does not necessarily rule out dynamic effects of past or future treatments (the latter in case of anticipation effects) on the outcome. The $K$ treatments may for instance include lags of the same treatment variables. We discuss this issue in more details after Theorem \ref{thm:extension} below, and in Web Appendix Section \ref{sub:single_treatment}.

\medskip
We consider the treatments and potential outcomes of each $(g,t)$ cell as random variables. For instance, aggregate random shocks may affect the potential outcomes of group $g$ at period $t$, and that cell's treatments may also be random. All expectations below are taken with respect to the distribution of those random variables. On the other hand, the populations of cells $(g,t)$ $N_{g,t}$ are treated as non-random throughout the paper.

\medskip
Throughout the paper, we maintain the following assumptions. Below, we let $\bm{0}=(0,...,0)$ denote the vector of $K$ zeros.

\begin{hyp}\label{hyp:supp_gt}
	(Balanced panel of groups) For all $(g,t)\in \{1,...,G\}\times  \{1,...,T\}$, $N_{g,t}>0$.
\end{hyp}
\begin{hyp}\label{hyp:independent_groups}
	(Independent groups) The vectors $((Y_{g,t}(\bm{d}))_{\bm{d}\in \{0,1\}^K},\bm{D}_{g,t})_{t\in \{1,...,T\}}$ are mutually independent.
\end{hyp}

\begin{hyp}\label{hyp:common_trends}
	(Strong exogeneity and common trends) For all $(g,t)\in \{1,...,G\}\times\{2,...,T\}$, \begin{enumerate}
\item $E(Y_{g,t}(\bm{0})-Y_{g,t-1}(\bm{0})|\bm{D}_{g,1},...,\bm{D}_{g,T})=E(Y_{g,t}(\bm{0})-Y_{g,t-1}(\bm{0})).$
\item $E(Y_{g,t}(\bm{0}) - Y_{g,t-1}(\bm{0}))$ does not vary across $g$.
\end{enumerate}
\end{hyp}

Assumption \ref{hyp:supp_gt} requires that no group appears or disappears over time. Assumption \ref{hyp:independent_groups} requires that potential outcomes and treatments of different groups be independent, but it allows these variables to be correlated over time  within each group. This is a commonly-made assumption in DID analysis, where standard errors are usually clustered at the group level \citep[see][]{bertrand2004}. Point 1 of Assumption \ref{hyp:common_trends} is related to the strong exogeneity condition in panel data models. It requires that the shocks affecting group $g$'s untreated outcome be mean independent of group $g$'s treatments. For instance, this rules out cases where a group gets treated because it experiences negative shocks, the so-called Ashenfelter's dip \citep[see][]{ashenfelter1978estimating}. Point 2 requires that in every group, the expectation of the untreated outcome follow the same evolution over time. It is a generalization of the standard common trends assumption in DID models \citep[see, e.g.,][]{Abadie05}.

\medskip
We now define the TWFE regression described in the introduction, as well as our estimand of interest $\beta_{fe}$, the expectation of the treatment coefficient in the regression.\footnote{\label{foot:well_def} Throughout the paper, we assume that the treatments $D^k_{g,t}$ in Regression \ref{reg1} are not collinear with the other independhent variables in those regressions, so $\widehat{\beta}_{fe}$ is well-defined.}
\begin{mydef}\label{reg1}
(TWFE regression with $K$ treatments)

\medskip
Let $\beta_{fe}=E\left[\widehat{\beta}_{fe}\right]$, where $\widehat{\beta}_{fe}$ denotes the coefficient on $D^1_{g,t}$ in a sample OLS regression of $Y_{g,t}$ on group fixed effects, period fixed effects, and the vector $\bm{D}_{g,t}$, weighted by $N_{g,t}$.\footnote{ \label{foot:disagg} The regression could also be estimated using more disaggregated outcome data. For instance, groups may be US counties, and one may estimate the regression using individual-level outcome measures. This disaggregated regression is equivalent to the aggregated regression, provided $Y_{g,t}$ is defined as the average outcome of individuals in cell $(g,t)$, and the aggregated regression is weighted by the number of individuals in cell $(g,t)$. Accordingly, the results below also apply to disaggregated regressions.}
\end{mydef}

On top of the $K$ treatments, the regression may also include some covariates. The decompositions below can easily be extended to this case, following the same steps as those used by \cite{dechaisemartin2020two} to extend their decomposition of TWFE regressions with one treatment to TWFE regressions with one treatment and some covariates (see Theorem S4 therein). 

\medskip
Let $\boldsymbol{D}$ be the vector $(\bm{D}_{g,t})_{(g,t)\in \{1,...,G\}\times \{1,...,T\}}$ collecting all the treatments in all the $(g,t)$ cells.
let $\bm{D}_g=(D_{1,g},...,D_{T,g})$ be the vector collecting all the treatments in group $g$.
Let $N_1=\sum_{g,t}N_{g,t}D^1_{g,t}$ denote the total population of cells receiving the first treatment. Let $\bm{D}^{-1}_{g,t}=(D^2_{g,t},...,D^K_{g,t})$ denote a vector stacking together the treatments of cell $(g,t)$, excluding treatment 1.
Let $\eps_{g,t}$ denote the residual of cell $(g,t)$ in the sample regression of $D^1_{g,t}$ on group and period fixed effects and $\bm{D}^{-1}_{g,t}$:
\begin{equation}
D^1_{g,t}=\widehat{\alpha}+\widehat{\gamma}_g+\widehat{\nu}_t+(\bm{D}^{-1}_{g,t})'\widehat{\bm{\zeta}} +\eps_{g,t}.
\label{eq:first_stage}
\end{equation}
If the regressors in Regression \ref{reg1} are not collinear, the average value of $\eps_{g,t}$ across all $(g,t)$ cells with $D^1_{g,t}=1$ differs from 0: $\sum_{(g,t):D^1_{g,t}=1}(N_{g,t}/N_1)\eps_{g,t} = \sum_{(g,t)}(N_{g,t}/N_1)\eps^2_{g,t}> 0$. Then we let $w_{g,t}$ denote $\eps_{g,t}$ divided by that average:
$$w_{g,t}=\frac{\eps_{g,t}}{\sum_{(g,t):D^1_{g,t}=1}(N_{g,t}/N_1)\eps_{g,t}}.$$

\section{TWFE regressions with several treatments}\label{sec:Identification}

\subsection{Decomposition results} 
\label{sub:decomp}

\subsubsection{Two treatment variables} 
\label{subsub:twotreatments}

For expositional purposes, we begin by considering the case with two treatments.
For any $(g,t)\in \{1,...,G\}\times \{1,...,T\}$, let $$\Delta^{2}_{g,t}=Y_{g,t}(0,1)-Y_{g,t}(0,0)$$
denote the effect, in cell $(g,t)$, of moving the second treatment from zero to 1 while keeping the first treatment at zero.
Let also
$$\Delta^{1}_{g,t}=Y_{g,t}(1,D^2_{g,t})-Y_{g,t}(0,D^2_{g,t})$$
denote the effect, in cell $(g,t)$, of moving the first treatment from zero to one while keeping the second treatment at its observed value.
When one estimates a TWFE regression with two treatments, a natural target parameter for $\beta_{fe}$, the coefficient on the first treatment, is
$$\delta_{ATT}=E\left[\sum_{(g,t):D^1_{g,t}=1}\frac{N_{g,t}}{N_1}\Delta^{1}_{g,t}\right],$$ the average effect of moving $D^1_{g,t}$ from 0 to 1 while keeping $D^2_{g,t}$ at its observed value, across all $(g,t)$s such that $D^1_{g,t}=1$. $\delta_{ATT}$ is the ATT of $D^1_{g,t}$ controlling for $D^2_{g,t}$. We now show that $\beta_{fe}$ does not identify $\delta_{ATT}$ in general.
\begin{thm}
Suppose that Assumptions \ref{hyp:supp_gt}-\ref{hyp:common_trends} hold and $K=2$.  Then,
\begin{align}\label{eq:thm:main}
\beta_{fe}  & =E\left[\sum_{(g,t):D^1_{g,t}=1}\frac{N_{g,t}}{N_1}w_{g,t}\Delta^{1}_{g,t}+\sum_{(g,t):D^2_{g,t}=1}\frac{N_{g,t}}{N_1}w_{g,t}\Delta^{2}_{g,t}\right].
\end{align}
Moreover, $\sum_{(g,t):D^1_{g,t}=1}(N_{g,t}/N_1)w_{g,t}=1$ and $\sum_{(g,t):D^2_{g,t}=1}(N_{g,t}/N_1)w_{g,t}=0$.
\label{thm:main}
\end{thm}
Theorem \ref{thm:main} shows that the coefficient on $D^1_{g,t}$ identifies the sum of two terms. The first term is a weighted sum of the average effect of moving $D^1_{g,t}$ from 0 to 1 while keeping $D^2_{g,t}$ at its observed value, across all $(g,t)$ such that $D^1_{g,t}=1$, and with weights summing to 1. The second term is a weighted sum of the effect of moving $D^2_{g,t}$ from 0 to 1 while keeping $D^1_{g,t}$ at 0, across all $(g,t)$ such that $D^2_{g,t}=1$, and with weights summing to 0. If the effect of $D^2_{g,t}$ is constant ($\Delta^{2}_{g,t}=\delta^2$ for all $(g,t)$), this second term is equal to zero, but it may differ from zero if the effect of $D^2_{g,t}$ is heterogeneous.

\medskip
Theorem \ref{thm:main} implies that there are two reasons why $\beta_{fe}$ may differ from $\delta_{ATT}$. First, some of the weights $w_{g,t}$ may differ from one. When the weights $w_{g,t}$  differ from one, one may have that $$E\left[\sum_{(g,t):D^1_{g,t}=1}\frac{N_{g,t}}{N_1}w_{g,t}\Delta^{1}_{g,t}\right]\ne \delta_{ATT},$$ if the effect of $D^1_{g,t}$ is heterogeneous across $(g,t)$ cells. Some of the weights $w_{g,t}$ could even be negative, in which case $E\left[\sum_{(g,t):D^1_{g,t}=1}(N_{g,t}/N_1) w_{g,t}\Delta^{1}_{g,t}\right]$  does not satisfy the no-sign reversal property: this quantity could for instance be negative, even if $\Delta^{1}_{g,t}\geq 0$ for all $(g,t)$. With two treatments, negative weights can occur even in very simple designs, where there would not be any negative weights in the absence of the second treatment. For instance, consider a standard DID set-up without variation in treatment timing but with two treatments: some groups start receiving the first treatment at a date $T^1$, and a subset of those groups then start receiving the second treatment at a later date $T^2$. In the absence of the second treatment, one can show that the coefficient on $D^1_{g,t}$ in the regression of $Y_{g,t}$ on group fixed effects, period fixed effects, and $D^1_{g,t}$ identifies the ATT of $D^1_{g,t}$ and does not have negative weights attached to it. On the other hand, in the presence of the second treatment, one can show that $\beta_{fe}$ no longer identifies the ATT of $D^1_{g,t}$ and may have negative weights attached to it \citep[see Corollary 1 in][a previous version of this paper, for a formal statement and a proof of these results]{DCDH_previous}.

\medskip
The second reason why $\beta_{fe}$ may differ from $\delta_{ATT}$ is that $\beta_{fe}$ may also be contaminated by the effect of $D^2_{g,t}$: if that effect is heterogeneous across $(g,t)$ cells, $E\left[\sum_{(g,t):D^2_{g,t}=1}(N_{g,t}/N_1)w_{g,t}\Delta^{2}_{g,t}\right]$ may differ from zero. Such a contamination phenomenon is not present in the presence of one treatment only \citep[see][]{dechaisemartin2020two}. Below, we give some intuition as to why it arises.

\medskip
Theorem \ref{thm:main} can be extended to non-binary ordered treatments, that may be continuous or discrete. When $D^1_{g,t}\ne 0$, let
$S^1_{g,t}=(Y_{g,t}(D^1_{g,t},D^2_{g,t})-Y_{g,t}(0,D^2_{g,t}))/D^1_{g,t}$ be the slope of cell $(g,t)$'s potential outcome function, when moving its first treatment from $0$ to $D^1_{g,t}$, while keeping its second treatment at its observed value. Similarly, when $D^2_{g,t}\ne 0$, let $S^2_{g,t}=(Y_{g,t}(0,D^2_{g,t})-Y_{g,t}(0,0))/D^2_{g,t}$. Finally, let
$$w^k_{g,t}=\frac{\eps_{g,t}D^k_{g,t}}{\sum_{(g,t)}(N_{g,t}/N_1)\eps_{g,t}D^1_{g,t}},$$
for $k=1, 2$. If $D^1_{g,t}$ and $D^2_{g,t}$ are non-binary, one can show, following similar steps as in the proof of Theorem \ref{thm:main}, that
$$\beta_{fe}  =E\left[\sum_{(g,t): D^1_{g,t}\ne 0}\frac{N_{g,t}}{N_1}w^1_{g,t}S^1_{g,t}+\sum_{(g,t): D^2_{g,t}\ne 0}\frac{N_{g,t}}{N_1}w^2_{g,t}S^2_{g,t}\right].$$
Moreover, $\sum_{(g,t): D^1_{g,t}\ne 0}(N_{g,t}/N_1)w^1_{g,t}=1$ and $\sum_{(g,t): D^2_{g,t}\ne 0} (N_{g,t}/N_1)w^2_{g,t}=0$.
Essentially, Theorem \ref{thm:main} extends to non-binary treatments, replacing the average treatment effects $\Delta^{1}_{g,t}$ and $\Delta^{2}_{g,t}$ by slopes of $(g,t)$-cells'  potential outcome functions, from a treatment of zero to their actual treatment. The decomposition in the previous display does not assume a linear treatment effect.

\subsubsection{More than two treatment variables} 
\label{subsub:extension}

We now go back to the general case where $K$ may be greater than 2. We let $\bm{0}^{-1}=(0,...,0)$ be the vector of $K-1$ zeros. We also define
\begin{align*}
\Delta^{1}_{g,t} = Y_{g,t}(1,\bm{D}^{-1}_{g,t}) - Y_{g,t}(0,\bm{D}^{-1}_{g,t}),\\	
\Delta^{-1}_{g,t} = Y_{g,t}(0,\bm{D}^{-1}_{g,t}) - Y_{g,t}(0,\bm{0}^{-1}).
\end{align*}
$\Delta^{1}_{g,t}$ is the effect, in cell $(g,t)$, of moving the first treatment from zero to one while keeping the other treatments at their observed values. $\Delta^{-1}_{g,t}$ is the effect, in cell $(g,t)$, of moving the other treatments from zero to their actual values, while keeping the first treatment at zero.

\medskip
Theorem \ref{thm:extension} below generalizes Theorem \ref{thm:main}.
\begin{thm}
Suppose that Assumptions \ref{hyp:supp_gt}-\ref{hyp:common_trends} hold.  Then,
\begin{align*}
\beta_{fe}  & =E\left[\sum_{(g,t):D^1_{g,t}=1}\frac{N_{g,t}}{N_1}w_{g,t}\Delta^{1}_{g,t}+\sum_{(g,t):\bm{D}^{-1}_{g,t}\ne \bm{0}^{-1}}\frac{N_{g,t}}{N_1}w_{g,t} \Delta^{-1}_{g,t} \right].
\end{align*}
Moreover, $\sum_{(g,t):D^1_{g,t}=1}(N_{g,t}/N_1)w_{g,t}=1$, and if $K=2$ or the treatments $D^2_{g,t},...,D^K_{g,t}$ are mutually exclusive, $\sum_{(g,t):\bm{D}^{-1}_{g,t}\ne \bm{0}^{-1}}(N_{g,t}/N_1)w_{g,t} = 0$.
\label{thm:extension}
\end{thm}
Theorem \ref{thm:extension} is similar to Theorem \ref{thm:main}, except that when $K>2$, we do not always have $$\sum_{(g,t):\bm{D}^{-1}_{g,t}\ne \bm{0}^{-1}}\frac{N_{g,t}}{N_1}w_{g,t}=0.$$
The contamination weights on the effects of the other treatments may not sum to 0. Accordingly, even if the effects of all treatments are constant, $\widehat{\beta}_{fe}$ may still be biased for the first treatment's effect.

\medskip
There are three special cases where the weights on the effects of the other treatments sum to 0. The first one is when $K=2$, as shown in Theorem \ref{thm:main}. The second one is when the treatments $D^2_{g,t},...,D^K_{g,t}$ are mutually exclusive, as stated in Theorem \ref{thm:extension}. The third one is when
there is no complementarity or substitutability between the treatments $D^2_{g,t},...,D^K_{g,t}$. Specifically, assume that for all $(g,t)$, there exists $(\delta^k_{g,t})_{k=2,...,K}$ such that
\begin{equation}
E\left[\Delta^{-1}_{g,t}|\bm{D}\right]  = \sum_{k=2}^K D^k_{g,t} \delta^k_{g,t}.	
	\label{eq:additive}
\end{equation}
Then, we obtain Decomposition \eqref{eq:decomp_add} below. The corresponding weights can be computed using the \st{twowayfeweights} Stata command.

\begin{cor}
Suppose that Assumptions \ref{hyp:supp_gt}-\ref{hyp:common_trends} and \eqref{eq:additive} hold.  Then,
\begin{equation}
\beta_{fe}   =E\left[\sum_{(g,t):D^1_{g,t}=1}\frac{N_{g,t}}{N_1}w_{g,t}\Delta^{1}_{g,t}+\sum_{k=2}^K\sum_{(g,t):D^k_{g,t}=1}\frac{N_{g,t}}{N_1}w_{g,t}\delta^k_{g,t}\right].
\label{eq:decomp_add}
\end{equation}
Moreover, $\sum_{(g,t):D^1_{g,t}=1}(N_{g,t}/N_1)w_{g,t}=1$, and $\sum_{(g,t):D^k_{g,t}=1}(N_{g,t}/N_1)w_{g,t}= 0$ for every $k\in \{2,...,K\}$.
\label{cor:additive}
\end{cor}

On the other hand, when the treatments are not mutually exclusive and may be complementary or substitutable $\widehat{\beta}_{fe}$ could be biased even under constant treatment effects. This is because in that case, Regression 1 is misspecified, and should include the interactions of the treatments.

\medskip
Importantly, Theorem \ref{thm:extension} does not necessarily rule out dynamic effects of past treatments on the outcome; similarly, it can allow for anticipation effects. The treatments in the regression may for instance be the current treatment, and some of its lags and leads. In that case, our potential outcome notation allows the current treatment and its lags and  leads included in the regression to affect the outcome. Accordingly, the \st{twowayfeweights} Stata command can also be used to compute the weights attached to distributed-lags regressions of an outcome on the current treatment, and some of its lags and leads.

\subsection{Intuition for, and a perhaps surprising implication of, the contamination bias} 
\label{sub:intuition_implications}

\subsubsection{Intuition for the contamination bias} 
\label{subsub:intuition}

The reasons why TWFE regressions are not robust to heterogeneous treatment effects are now well understood \citep[see][]{deChaisemartin15b,dechaisemartin2020two,goodman2021difference,borusyak2016}. In this section, we give intuition as to why $\beta_{fe}$ may be affected by contamination bias. To do so, we start by considering two very simple examples, one where contamination bias is absent, and the other where it is present.

\medskip
First, assume that there are three groups and two time periods. With probability one, no group is treated at period 1, and at period 2 group 2 receives the first treatment while group 3 receives the second treatment. Then, it is easy to show that
\begin{equation}\label{eq:example1DID}
\widehat{\beta}_{fe}=Y_{2,2}-Y_{2,1}-\left(Y_{1,2}-Y_{1,1}\right).
\end{equation}
The right-hand side of the previous display is a DID comparing the period-one-to-two outcome evolution of group 2, that starts receiving the first treatment at period 2, to that of group 1, that is untreated at both dates.
Therefore,
\begin{align}\label{eq:example1}
\beta_{fe}=&E\left(Y_{2,2}(1,0)-Y_{2,1}(0,0)-\left(Y_{1,2}(0,0)-Y_{1,1}(0,0)\right)\right)\nonumber \\
=&E\left(Y_{2,2}(1,0)-Y_{2,2}(0,0)\right)+E\left(Y_{2,2}(0,0)-Y_{2,1}(0,0)-\left(Y_{1,2}(0,0)-Y_{1,1}(0,0)\right)\right)\nonumber\\ =&E\left(Y_{2,2}(1,0)-Y_{2,2}(0,0)\right),
\end{align}
where the second equality follows from Assumption \ref{hyp:common_trends}. Equation \eqref{eq:example1} is a special case of Equation \eqref{eq:thm:main} in Theorem \ref{thm:main}. In this simple example, $\beta_{fe}$ is not contaminated by the effect of the second treatment. It identifies the effect, in group 2 and at period 2, of moving the first treatment from zero to one while keeping the second treatment at its observed value (zero). Because only group 2 at period 2 receives the first treatment, this effect is equal to $\delta_{ATT}$, the ATT of the first treatment controlling for the second treatment.

\medskip
Now let us consider another example, very similar to that above, but with a fourth group that receives both treatments at period 2. Then, using the equivalence between TWFE regressions and first-difference regressions with two periods and the fact that the first difference of the two treatments are uncorrelated, we obtain
\begin{equation}\label{eq:example2DID}
\widehat{\beta}_{fe}=\frac{1}{2}\left(Y_{2,2}-Y_{2,1}-\left(Y_{1,2}-Y_{1,1}\right)\right)+\frac{1}{2}\left(Y_{4,2}-Y_{4,1}-\left(Y_{3,2}-Y_{3,1}\right)\right).
\end{equation}
The first DID in Equation \eqref{eq:example2DID} is the same as that in the right-hand side of Equation \eqref{eq:example1DID} and it is unbiased for $E\left(Y_{2,2}(1,0)-Y_{2,2}(0,0)\right)$. The second DID compares the period-one-to-two outcome evolution of group 4, that starts receiving the first and second treatments at period 2, to that of group 3, that only starts receiving the second treatment.
Therefore,
\begin{align}\label{eq:example2DID2}
&E\left(Y_{4,2}-Y_{4,1}-\left(Y_{3,2}-Y_{3,1}\right)\right)\nonumber \\
=&E\left(Y_{4,2}(1,1)-Y_{4,1}(0,0)-\left(Y_{3,2}(0,1)-Y_{3,1}(0,0)\right)\right)\nonumber \\
=&E\left(Y_{4,2}(1,1)-Y_{4,2}(0,1)\right)+E\left(Y_{4,2}(0,1)-Y_{4,2}(0,0)\right)-E\left(Y_{3,2}(0,1)-Y_{3,2}(0,0)\right)\nonumber \\
+&E\left(Y_{4,2}(0,0)-Y_{4,1}(0,0)-\left(Y_{3,2}(0,0)-Y_{3,1}(0,0)\right)\right)\nonumber\\
=&E\left(Y_{4,2}(1,1)-Y_{4,2}(0,1)\right)+E\left(Y_{4,2}(0,1)-Y_{4,2}(0,0)\right)-E\left(Y_{3,2}(0,1)-Y_{3,2}(0,0)\right).
\end{align}
Equations \eqref{eq:example2DID} and \eqref{eq:example2DID2} imply that
\begin{align}\label{eq:example2}
\beta_{fe}=&\frac{1}{2}E\left(Y_{2,2}(1,0)-Y_{2,2}(0,0)\right)+\frac{1}{2}E\left(Y_{4,2}(1,1)-Y_{4,2}(0,1)\right)\nonumber\\
+&\frac{1}{2}E\left(Y_{4,2}(0,1)-Y_{4,2}(0,0)\right)-\frac{1}{2}E\left(Y_{3,2}(0,1)-Y_{3,2}(0,0)\right).
\end{align}
Equation \eqref{eq:example2} is a special case of Equation \eqref{eq:thm:main} in Theorem \ref{thm:main}.
$\beta_{fe}$ identifies the sum of two terms. The term on the first line is the average effect, in groups two and four and at period two, of moving the first treatment from zero to one while keeping the second treatment at its observed value (zero in group 2, one in group 4). The term on the second line is a contamination bias term, equal to the difference, between groups 4 and 3, of the effect of moving the second treatment from zero to one while keeping the first treatment at zero.

\medskip
The contamination bias appears in the second example because $\widehat{\beta}_{fe}$ leverages a DID comparing a group that starts receiving the first and the second treatments to a group that starts receiving the second treatment only. With heterogeneous treatment effects, this comparison is contaminated by the effect of the second treatment. On the other hand, if the effect of the second treatment does not vary across groups, this contamination bias disappears. To our knowledge, our paper is the first to show that TWFE regressions with several treatments leverage this type of ``forbidden comparisons'', using the terminology coined by \cite{borusyak2016}.

\medskip
In the example with four groups, a simple solution to eliminate the contamination bias is to add the interaction of the two treatments to the regression. One can in fact show the following, slightly more general result. With only two time periods, and groups that do not receive any of the two treatments in the first period, the coefficient on $D^1_{g,t}$ in the regression of $Y_{g,t}$ on $D^1_{g,t}$, $D^2_{g,t}$, and $D^1_{g,t}D^2_{g,t}$ is not contaminated by the effect of the second treatment. In such cases, the regression with the interaction term is preferable, as it makes the contamination problem disappear. This result does not, however, translate to more general designs with more than two time periods and where groups may receive the treatments at every period.  It is easy to find examples where adding the interaction to the regression actually increases the contamination weights. This is the case for instance in the application we consider in Section \ref{sec:appli}: in the regression without control variables and with the two main treatments (the minimum staff-to-child ratio and the minimum number of years of schooling required for daycare directors), adding the interaction between the two treatments actually increases the absolute value of the contamination weights.

\medskip
In the first example, the two treatments are mutually exclusive so $\widehat{\beta}_{fe}$ cannot leverage a ``forbidden'' DID comparing a group that starts receiving the first and the second treatments to a group that starts receiving the second treatment only, which is why there is no contamination bias in this example. This does not mean contamination bias never arises with mutually exclusive treatments. To illustrate this point, let us consider a third example with two groups and three periods. Group 1 receives the first treatment at period 3, and Group 2 receives the second treatment at periods 2 and 3. Then, because this regression is equivalent to a regression of $Y_{2,t}-Y_{1,t}$ on a constant, $D^1_{2,t}-D^1_{1,t}$ and $D^2_{2,t}-D^2_{1,t}$, we obtain, after some algebra,
\begin{equation}\label{eq:example3DID}
\widehat{\beta}_{fe}=Y_{1,3}-Y_{1,2}-\left(Y_{2,3}-Y_{2,2}\right).
\end{equation}
Accordingly, one can show that
\begin{align}\label{eq:example3}
\beta_{fe}=&E\left(Y_{1,3}(1,0)-Y_{1,3}(0,0)\right)\nonumber\\
+&E\left(Y_{2,2}(0,1)-Y_{2,2}(0,0)\right)-E\left(Y_{2,3}(0,1)-Y_{2,3}(0,0)\right).
\end{align}
$\widehat{\beta}_{fe}$ is contaminated by the effect of the second treatment, because it leverages a DID where the control group receives the second treatment at both dates. This second type of ``forbidden'' DID is very similar to the late- versus early-treated DIDs due to which TWFE regressions with one treatment are not robust to heterogeneous treatment effects \citep[see][]{dechaisemartin2020two,goodman2021difference,borusyak2016}. Note that if the effect of the second treatment is constant over time, the contamination bias term disappears. However, constant effects over time is often an implausible assumption.

\medskip
Overall, TWFE regressions with several treatments are not affected by contamination bias in very simple designs with two time periods, where groups are only treated in the second period, and where the treatments are mutually exclusive. In designs with non-mutually exclusive treatments, contamination bias may appear because $\widehat{\beta}_{fe}$ may leverage DIDs comparing a group that starts receiving, say, the first and the second treatments to a group that starts receiving the second treatment only. With more than two time periods, even if the treatments are mutually exclusive, $\widehat{\beta}_{fe}$ may leverage DIDs comparing a group that starts receiving, say, the first treatment, to a group receiving the second treatment at both dates.

\subsubsection{A perhaps surprising implication of the contamination bias} 
\label{subsub:implications}

Theorem \ref{thm:main} has an important and perhaps surprising consequence for TWFE regressions with one treatment where one seeks to estimate heterogeneous treatment effects. Oftentimes, researchers run a TWFE regression with a treatment variable $D_{g,t}$ interacted with a group-level binary variable $I_g$, and with $(1-I_g)$.\footnote{Researchers may instead have $D_{g,t}$ and $D_{g,t}I_g$ in the regression. The coefficient on $D_{g,t}$ in this regression is equal to that on $D_{g,t}(1-I_g)$ in the regression described in the text. The coefficient on $D_{g,t}I_g$ is equal to the difference between that on $D_{g,t}I_g$ and that on $D_{g,t}(1-I_g)$ in the regression described in the text. Accordingly, the discussion in this section also applies to those regressions.} For instance, to study if the treatment effect differs in poor and rich counties, one interacts the treatment with an indicator for counties above the median income, and with an indicator for counties below the median income. Theorem \ref{thm:main} also applies to those regressions. Specifically, one has
\begin{align*}
\beta^{I=1}_{fe}  & =E\left[\sum_{(g,t):D_{g,t}=1,I_g=1}\frac{N_{g,t}}{N_1}w_{g,t}\Delta_{g,t}+\sum_{(g,t):D_{g,t}=1,I_g=0}\frac{N_{g,t}}{N_1}w_{g,t}\Delta_{g,t}\right].
\end{align*}
where $\beta^{I=1}_{fe}$ is the coefficient on $D_{g,t}\times I_g$, and
$\Delta_{g,t}=Y_{g,t}(1)-Y_{g,t}(0).$
The previous display implies that the coefficient on $D_{g,t}\times I_g$ is contaminated by the treatment effect in $(g,t)$ cells such that $I_g=0$. In the example, the coefficient on the treatment interacted with the indicator for rich counties is contaminated by the treatment effect in poor counties. This calls into question the use of such TWFE regressions to estimate heterogeneous effects.

\medskip
This contamination phenomenon disappears if the time fixed effects are interacted with $I_g$ in the regression. Then, the coefficient on $D_{g,t}\times I_g$ becomes equivalent to that one would obtain by running a TWFE regression restricting the sample to groups such that $I_g=1$. It follows from \cite{dechaisemartin2020two} that this coefficient identifies a weighted sum of the treatment effects across $(g,t)$ cells such that $D_{g,t}=1,I_g=1$: it is not contaminated by the treatment effect in $(g,t)$ cells such that $D_{g,t}=1,I_g=0$.

\subsection{Should one control for other treatments?} 
\label{sub:OVB}

In this section, we derive a decomposition similar to that in Theorem \ref{thm:main}, when there are two treatments but the second treatment is omitted from the regression.
\begin{mydef}\label{regshort}
(Short TWFE regression)

\medskip
Let $\beta^s_{fe}=E\left[\widehat{\beta}^s_{fe}\right]$, where $\widehat{\beta}^s_{fe}$ denotes the coefficient on $D^1_{g,t}$ in a sample OLS regression of $Y_{g,t}$ on group fixed effects, period fixed effects, and $D^1_{g,t}$, weighted by $N_{g,t}$.
\end{mydef}
Let $\eps^s_{g,t}$ denote the residual of cell $(g,t)$ in the sample regression of $D^1_{g,t}$ on group and period fixed effects.
If the regressors in Regression \ref{regshort} are not collinear, the average value of $\eps^s_{g,t}$ across all $(g,t)$ cells with $D^1_{g,t}=1$ differs from 0: $\sum_{(g,t):D^1_{g,t}=1}(N_{g,t}/N_1)\eps^s_{g,t}\neq 0$. Then we let $w^s_{g,t}$ denote $\eps^s_{g,t}$ divided by that average:
$$w^s_{g,t}=\frac{\eps^s_{g,t}}{\sum_{(g,t):D^1_{g,t}=1}(N_{g,t}/N_1)\eps^s_{g,t}}.$$
\begin{thm}
Suppose that Assumptions \ref{hyp:supp_gt}-\ref{hyp:common_trends} hold and $K=2$.  Then,
\begin{align}\label{eq:thm:OVB}
\beta^s_{fe}  & =E\left[\sum_{(g,t):D^1_{g,t}=1}\frac{N_{g,t}}{N_1}w^s_{g,t}\Delta^{1}_{g,t}+\sum_{(g,t):D^2_{g,t}=1}\frac{N_{g,t}}{N_1}w^s_{g,t}\Delta^{2}_{g,t}\right].
\end{align}
Moreover, $\sum_{(g,t):D^1_{g,t}=1}(N_{g,t}/N_1)w^s_{g,t}=1$ and $\sum_{(g,t):D^2_{g,t}=1}(N_{g,t}/N_1)w^s_{g,t}$ may differ from zero.
\label{thm:OVB}
\end{thm}
Theorem \ref{thm:OVB} is similar to Theorem \ref{thm:main}. It shows that the coefficient on $D^1_{g,t}$ in the short regression identifies the sum of two terms. The first term in Theorem \ref{thm:OVB} is similar to that in Theorem \ref{thm:main}, namely a weighted sum of the effect of moving $D^1_{g,t}$ from 0 to 1 while keeping $D^2_{g,t}$ at its observed value, but with different weights that still sum to one. The second term in Theorem \ref{thm:OVB} is also similar to that in Theorem \ref{thm:main}, namely a weighted sum of the effect of moving $D^2_{g,t}$ from 0 to 1 while keeping $D^1_{g,t}$ at 0, but with different weights that no longer sum to zero. Note that Theorem \ref{thm:OVB} can easily be extended to instances with more than two treatments.

\medskip
Theorem \ref{thm:OVB} is also similar to Theorem 1 in \cite{dechaisemartin2020two}. With the notation of this paper, Theorem 1 in \cite{dechaisemartin2020two} provides a decomposition of $\beta^s_{fe}$ under a parallel trends assumption on $Y_{g,t}(0,D^{-1}_{g,t})$, the potential outcome of $g$ at $t$ with the first treatment set at $0$ and the other treatments set at their actual values. The weighted sum of effects in Theorem 1 of \cite{dechaisemartin2020two} is identical to the first weighted sum in Theorem \ref{thm:OVB}. On the other hand, the contamination term in Theorem \ref{thm:OVB} does not appear in Theorem 1 of \cite{dechaisemartin2020two}, because the parallel trend assumptions underlying the two results are not the same. There may be instances where parallel trends on $Y_{g,t}(0,D^{-1}_{g,t})$ is plausible, in which case the decomposition in Theorem 1 in \cite{dechaisemartin2020two} is applicable. In this paper, as the researcher estimates a TWFE regression with several treatments, it is natural to consider instead a parallel trends assumption on $Y_{g,t}(\bm{0})$.

\medskip
Under constant effects and a parallel trends assumption on $Y_{g,t}(0,0)$, omitting the second treatment from the regression leads to an omitted variable bias, and including the second treatment into the regression is always preferable. But this may not be the case with heterogeneous treatment effects: because the weights associated with the two regressions differ, $D^1_{g,t}$'s coefficient in the long regression may be more biased for $\delta_{ATT}$ than $D^1_{g,t}$'s coefficient in the short regression. The following corollary formalizes this idea.

\begin{cor}
Suppose that Assumptions \ref{hyp:supp_gt}-\ref{hyp:common_trends} hold, $K=2$, and there is a real number $B$ such that $|\Delta^{1}_{g,t}|\leq B$ and $|\Delta^{2}_{g,t}|\leq B$ for all $(g,t)$. Then,
\begin{align*}
|\beta_{fe}-\delta_{ATT}|  & \leq B \times E\left[\sum_{(g,t):D^1_{g,t}=1}\frac{N_{g,t}}{N_1}|w_{g,t}-1|+\sum_{(g,t):D^2_{g,t}=1}\frac{N_{g,t}}{N_1}|w_{g,t}|\right],\\
|\beta^s_{fe}-\delta_{ATT}|  & \leq B \times E\left[\sum_{(g,t):D^1_{g,t}=1}\frac{N_{g,t}}{N_1}|w^s_{g,t}-1|+\sum_{(g,t):D^2_{g,t}=1}\frac{N_{g,t}}{N_1}|w^s_{g,t}|\right].	
\end{align*}
Moreover, both upper bounds are sharp.
\label{cor:maxbiasshortlong}
\end{cor}
Corollary \ref{cor:maxbiasshortlong} assumes that the effects of the first and second treatments are both bounded in every $(g,t)$ cell by a constant $B$. Under that assumption, it gives the maximal
biases of $\widehat{\beta}_{fe}$ and $\widehat{\beta}^s_{fe}$ as estimators of $\delta_{ATT}$, the ATT of $D^1_{g,t}$ controlling for $D^2_{g,t}$. One can compare those maximal biases by comparing (estimates of) $$E\left[\sum_{(g,t):D^1_{g,t}=1}\frac{N_{g,t}}{N_1}|w_{g,t}-1|+\sum_{(g,t):D^2_{g,t}=1}\frac{N_{g,t}}{N_1}|w_{g,t}|\right]$$
and $$E\left[\sum_{(g,t):D^1_{g,t}=1}\frac{N_{g,t}}{N_1}|w^s_{g,t}-1|+\sum_{(g,t):D^2_{g,t}=1}\frac{N_{g,t}}{N_1}|w^s_{g,t}|\right],$$ which does not require specifying $B$.\footnote{A similar result holds if we consider distinct bounds $B_1$ and $B_2$ for $|\Delta^{1}_{g,t}|$ and $|\Delta^{2}_{g,t}|$. Then, one has to multiply $ \sum_{(g,t):D^2_{g,t}=1}(N_{g,t}/N_1)|w_{g,t}|$ by $(B_2/B_1)$ when performing the comparison of the maximal biases. Hence, in this case, one needs to take a stand on the ratio $B_2/B_1$.} The maximal
bias of  $\widehat{\beta}_{fe}$ could be larger than that of $\widehat{\beta}^s_{fe}$, if for $(g,t)$s such that $D^1_{g,t}=1$ the weights $w_{g,t}$ are on average further away from one than the weights $w^s_{g,t}$, and/or if for $(g,t)$s such that $D^2_{g,t}=1$ the contamination weights $w_{g,t}$ are on average further away from zero than the weights $w^s_{g,t}$.
In our application in Section \ref{sec:appli}, we find that the estimated maximal bias of the long regression is almost five times larger than that of the short regression. Then, the short regression is preferable, at least per our maximal-bias
metric.

\section{Alternative estimator}\label{sec:alternative_estimand} 

\subsection{Identifying assumption}\label{sub:alternative_idassumption}

In this section, we start by considering the following identifying assumption. Recall that $Y_{g,t}(\bm{d})$ denotes the potential outcome of $g$ at $t$, if the treatment vector is equal to $\bm{d}$.

\begin{hyp}\label{hyp:CT_alt}
	(Strong exogeneity and common trends from $t-1$ to $t$, conditional on $\bm{D}_{g,t-1}$)
For all $(g,t)\in \{1,...,G\}\times\{2,...,T\}$ and all $\bm{d}\in \{0,1\}^K$,
\begin{enumerate}
\item $E(Y_{g,t}(\bm{d})-Y_{g,t-1}(\bm{d})|\bm{D}_{g,1},...,\bm{D}_{g,t-2},\bm{D}_{g,t-1}=\bm{d},\bm{D}_{g,t},...,\bm{D}_{g,T})=E(Y_{g,t}(\bm{d})-Y_{g,t-1}(\bm{d})|\bm{D}_{g,t-1}=\bm{d}).$
\item $E(Y_{g,t}(\bm{d})-Y_{g,t-1}(\bm{d})|\bm{D}_{g,t-1}=\bm{d})$ does not vary across $g$.
\end{enumerate}
\end{hyp}
Like Assumption \ref{hyp:common_trends}, Assumption \ref{hyp:CT_alt} imposes both a strong exogeneity and a parallel trends condition. The strong exogeneity condition requires that groups' $t-1$-to-$t$ outcome evolution, in the counterfactual scenario where their period-$t$ treatments all remain at their $t-1$ value, be mean independent of their treatments at every period other than $t-1$. The parallel trends assumption requires that groups with the same period-$t-1$ treatments have the same counterfactual trends. Then, consider a group whose first treatment changes between $t-1$ and $t$, but whose other treatments remain constant. Under Assumption \ref{hyp:CT_alt}, the $t-1$-to-$t$ evolution of its outcome had its first treatment not changed is identified by the outcome evolution of groups whose treatments all remain constant and with the same period-$t-1$ treatments.

\medskip
We now compare our new assumption, Assumption \ref{hyp:CT_alt}, to the more standard Assumption \ref{hyp:common_trends}. The two assumptions are non-nested, and there are two main differences between them.
First, Assumption \ref{hyp:common_trends} requires that all groups be on parallel trends, over the entire duration of the panel. Assumption \ref{hyp:CT_alt}, on the other hand, only requires that groups with the same period-$t-1$  treatments be on parallel trends, from $t-1$ to $t$. Assumption \ref{hyp:CT_alt} may then be more plausible: groups with the same treatments in the baseline period may be more similar, and may be more likely to experience parallel trends.\footnote{Because it imposes parallel trends conditional on $\bm{D}_{g,t-1}$, Assumption \ref{hyp:CT_alt} may be seen as ``in-between'' a standard parallel trends assumption and the sequential ignorability assumption, another commonly-used identifying assumption in panel data models \citep[see, e.g.,][]{robins1986new,bojinov2020panel}. Sequential ignorability requires that treatment be uncounfounded conditional on prior treatment and outcome, which implies parallel trends conditional on prior treatment and outcome. Because Assumption \ref{hyp:CT_alt} does not condition on groups' $t-1$ outcomes, it may be less plausible than sequential ignorability. At the same time, estimators relying on sequential ignorability need to compare groups with the same prior treatments and outcomes. This may lead to a curse of dimensionality.} Moreover, parallel trends may be more likely to hold over consecutive time periods than over the panel's entire duration.

\medskip
Second, Assumption \ref{hyp:common_trends} is a parallel trends assumption in the counterfactual where groups do not receive any treatment, while Assumption \ref{hyp:CT_alt} is a parallel trends assumption in the counterfactual where groups' treatments do not change from $t-1$ to $t$. Accordingly, Assumption \ref{hyp:common_trends} only restricts one potential outcome, the one without any treatment, while Assumption \ref{hyp:CT_alt} imposes restrictions on many potential outcomes. Still, Assumption \ref{hyp:CT_alt} does not impose any restriction on treatment effect heterogeneity, because it restricts only one potential outcome per $(g,t)$ cell, namely $Y_{g,t}(\bm{d})$ for $(g,t)$ cells such that $\bm{D}_{g,t-1}=\bm{d}$. In particular, Assumption \ref{hyp:CT_alt} does not require that all groups experience the same evolution of their treatment effect. Moreover, in complicated designs where the number of treatments is large and/or when the treatments are non binary, Assumption \ref{hyp:CT_alt} may have considerably more identifying power than Assumption \ref{hyp:common_trends}. Under Assumption \ref{hyp:common_trends}, an heterogeneity-robust DID estimator can only use as controls groups that do not receive any treatment at two dates at least. Moreover, treatment effects can only be estimated for groups that do not receive any treatment at one date at least. With many treatments and/or when the treatments are non binary, those two sets of groups may be small. In our empirical application in Section \ref{sec:appli}, there are two non-binary treatments, and while there are $(g,t)$ cells whose two treatments are equal to 0, there is no group that does not receive any of the two treatments at two dates at least. Accordingly, we cannot construct an heterogeneity-robust DID estimator relying on Assumption \ref{hyp:common_trends}, while we can construct one relying on Assumption \ref{hyp:CT_alt}.

\medskip
Importantly, there are special cases where: i) the two assumptions are equivalent and ii) our decomposition of $\widehat{\beta}_{fe}$ under Assumption \ref{hyp:common_trends} has contamination weights attached to it. This shows that decomposing $\widehat{\beta}_{fe}$ under Assumption \ref{hyp:CT_alt} can also lead to contamination weights. For instance, with two periods and $\bm{D}_{g,1}=0$ almost surely, Assumption \ref{hyp:common_trends} and \ref{hyp:CT_alt} are equivalent.  As a result, contamination weights may arise under Assumption \ref{hyp:CT_alt}, as the example we give p.13 demonstrates. In designs where Assumptions \ref{hyp:CT_alt} and \ref{hyp:common_trends} are not equivalent, under Assumption \ref{hyp:CT_alt} we cannot in general write $\beta_{fe}$ as a function of the design and treatment effects only, and replacing Assumption \ref{hyp:common_trends} by Assumption \ref{hyp:CT_alt} may actually exacerbate the problems of TWFE regressions: in addition to not being robust to heterogeneous treatment effects, TWFE regressions may now be biased even with homogenous treatment effects. We provide an example in Web Appendix Section \ref{appendix:decompo_altCT}.

\medskip
We also consider a second identifying assumption.
\begin{hyp}\label{hyp:CT_alt2}
	(Strong exogeneity and common trends from $t-1$ to $t$, conditional on $\bm{D}_{g,t}$)
For all $(g,t)\in \{1,...,G\}\times\{2,...,T\}$ and all $\bm{d}_{t}\in \{0,1\}^K$,
\begin{enumerate}
\item $E(Y_{g,t}(\bm{d}_{t})-Y_{g,t-1}(\bm{d}_{t})|\bm{D}_{g,1},...,\bm{D}_{g,t-1},\bm{D}_{g,t}=\bm{d}_{t},\bm{D}_{g,t+1},...,\bm{D}_{g,T})=E(Y_{g,t}(\bm{d}_{t})-Y_{g,t-1}(\bm{d}_{t})$ $|\bm{D}_{g,t}=\bm{d}_{t}).$
\item $E(Y_{g,t}(\bm{d}_{t})-Y_{g,t-1}(\bm{d}_{t})|\bm{D}_{g,t}=\bm{d}_{t})$ does not vary across $g$.
\end{enumerate}
\end{hyp}
Assumption \ref{hyp:CT_alt2} is similar to Assumption \ref{hyp:CT_alt}, except that it assumes parallel trends from $t-1$ to $t$, in the counterfactual where groups keep their period-$t$ rather than their period-$t-1$ treatments. Imposing jointly Assumptions \ref{hyp:CT_alt} and \ref{hyp:CT_alt2} may imply that the treatment effects follow the same evolution over time in some groups.\footnote{For instance, if $K=1$, $G=4$, $T=2$, $D^1_{1,1}=D^1_{1,2}=0$, $D^1_{2,1}=D^1_{2,2}=1$, $D^1_{3,1}=0,D^1_{3,2}=1$, and
$D^1_{4,1}=1, D^1_{4,2}=0$, one can show that together, Assumptions \ref{hyp:CT_alt} and \ref{hyp:CT_alt2} imply that the treatment effect follows the same evolution in groups $3$ and $4$.}

\subsection{Target parameters}\label{sub:alternative_target}

Let us define
\begin{align*}
\mathcal{S}_1=&\bigg\{(g,t): t\geq 2,D^1_{g,t}\ne D^1_{g,t-1},\bm{D}^{-1}_{g,t}=\bm{D}^{-1}_{g,t-1}, \exists g':
\bm{D}_{g',t} = \bm{D}_{g',t-1}=\bm{D}_{g,t-1}\bigg\}
\end{align*}
and let $N_{\mathcal{S}_1}=\sum_{(g,t)\in\mathcal{S}_1}N_{g,t}$. $\mathcal{S}_1$ is the set of cells $(g,t)$ whose first treatment changes between $t-1$ and $t$ while their other treatments do not change, and such that there is another group $g'$ whose treatments do not change between $t-1$ and $t$, and with the same treatments as $g$ in $t-1$. Hereafter, those cells are referred to as switchers. We show below that under Assumption \ref{hyp:CT_alt}, one can unbiasedly estimate
\begin{align*}
\delta_{1}&=E\left[\sum_{(g,t)\in \mathcal{S}_1} \frac{N_{g,t}}{N_{\mathcal{S}_1}} \Delta^{1}_{g,t}\right],
\end{align*}
the average effect of moving the first treatment from 0 to 1 while keeping all other treatments at their observed value, across all switchers.\footnote{When $N_{\mathcal{S}_1}=0$, we simply let the term inside brackets be equal to 0.}

\medskip
$\delta_{1}$ may differ from $\delta_{ATT}$, arguably a more natural target parameter. The two parameters apply to different and non-nested sets of $(g,t)$ cells. Let $\mathcal{T}_1=\{(g,t):D^1_{g,t}=1\}$. $\delta_{1}$ is the average of $\Delta^{1}_{g,t}$ across all cells in $\mathcal{S}_1$. $\delta_{ATT}$ is the average effect of $\Delta^{1}_{g,t}$ across all cells in $\mathcal{T}_1$. The following proposition shows that in our set-up, we cannot identify treatment effects on cells outside $\mathcal{S}_1$.

\begin{prop}
	Suppose that Assumptions \ref{hyp:supp_gt}-\ref{hyp:independent_groups} and \ref{hyp:CT_alt} hold. Then, for any subset $\mathcal{V}$ of $\mathcal{T}_1\backslash \mathcal{S}_1$, and letting $N_{\mathcal{V}}=\sum_{(g,t)\in\mathcal{V}} N_{g,t}$, $E[\sum_{(g,t)\in \mathcal{V}} (N_{g,t}/N_{\mathcal{V}}) \Delta^{1}_{g,t}]$ is not identified.
	\label{prop:maximal_set}
\end{prop}

Proposition \ref{prop:maximal_set} shows that $\mathcal{S}_1$ is the maximal set of cells for which treatment effects can be identified under Assumptions \ref{hyp:supp_gt}-\ref{hyp:independent_groups} and \ref{hyp:CT_alt}. The $(g,t)$ cells belonging to $\mathcal{T}_1$ but not to $\mathcal{S}_1$ can be divided into five mutually exclusive subgroups, detailed in Web Appendix Section \ref{appendix:diffATTLATE}. Identifying the effect of the first treatment in each of those subgroups would either require restricting treatment effect heterogeneity, or making parallel trend restrictions different from those in Assumption \ref{hyp:CT_alt}.

\medskip
While we expect  $\mathcal{S}_1$ to be often smaller than $\mathcal{T}_1$, there are also $(g,t)$ cells that belong to $\mathcal{S}_1$ but not to $\mathcal{T}_1$. Those are the switching-out cells, such that $D^1_{g,t}=0,D^1_{g,t-1}=1,\bm{D}^{-1}_{g,t}=\bm{D}^{-1}_{g,t-1}, \exists g':
\bm{D}_{g',t} = \bm{D}_{g',t-1}=\bm{D}_{g,t-1}$.

\medskip
As $\delta_{1}$ and  $\delta_{ATT}$ apply to different, non-nested subpopulations, a significant difference between $\widehat{\beta}_{fe}$ and the estimator of $\delta_{1}$ we propose below cannot be interpreted as evidence that $\widehat{\beta}_{fe}$ is biased for $\delta_{ATT}$. It could also be the case that $\widehat{\beta}_{fe}$ is unbiased for $\delta_{ATT}$ and $\delta_{1}$ and $\delta_{ATT}$ differ. On the other hand, under Assumptions \ref{hyp:common_trends} and \ref{hyp:CT_alt}, a significant difference  between $\widehat{\beta}_{fe}$ and the estimator of $\delta_{1}$ implies that the effect of at least one treatment is not constant.

\medskip
Similarly, we show below that under Assumption \ref{hyp:CT_alt2}, one can unbiasedly estimate
\begin{align*}
\delta_{2}&=E\left[\sum_{(g,t)\in \mathcal{S}_2} \frac{N_{g,t}}{N_{\mathcal{S}_2}} \Delta^{1}_{g,t}\right],
\end{align*}
where
\begin{align*}
\mathcal{S}_2=&\bigg\{(g,t): t\leq T-1,D^1_{g,t}\ne D^1_{g,t+1},\bm{D}^{-1}_{g,t}=\bm{D}^{-1}_{g,t+1}, \exists g':
\bm{D}_{g',t} = \bm{D}_{g',t+1}=\bm{D}_{g,t+1}\bigg\},
\end{align*}
and $N_{\mathcal{S}_2}=\sum_{(g,t)\in\mathcal{S}_2}N_{g,t}$. $\mathcal{S}_2$ is the set of cells $(g,t)$ whose first treatment changes between $t$ and $t+1$ while their other treatments do not change, and such that there is another group $g'$ whose treatments do not change between $t$ and $t+1$, and with the same treatments as $g$ in $t+1$. $\mathcal{S}_1$ and $\mathcal{S}_2$ are not necessarily disjoints: a $(g,t)$ cell experiencing two consecutive changes of its first treatment ($D^1_{g,t-1}\ne D^1_{g,t}$ and $D^1_{g,t}\ne D^1_{g,t+1}$) may belong both to $\delta_{1}$ and to $\delta_{2}$. On the other hand, a $(g,t)$ cell that does not experience two consecutive changes of its first treatment ($D^1_{g,t-1}=D^1_{g,t}$ or $D^1_{g,t}=D^1_{g,t+1}$) may belong to $\delta_{1}$ or to $\delta_{2}$ but cannot belong to both sets.

\medskip
Finally, under Assumptions \ref{hyp:CT_alt} and \ref{hyp:CT_alt2}, one can unbiasedly estimate
\begin{align*}
\delta_{}&=E\left[\sum_{(g,t)\in \mathcal{S}_1 \cup \mathcal{S}_2} \frac{N_{g,t}}{N_{\mathcal{S}_1 \cup \mathcal{S}_2}} \Delta^{1}_{g,t}\right],
\end{align*}
where $N_{\mathcal{S}_1 \cup \mathcal{S}_2}=\sum_{(g,t)\in \mathcal{S}_1 \cup \mathcal{S}_2}N_{g,t}$.

\subsection{Estimation}\label{sub:alternative_estimator}

We now show that under Assumption \ref{hyp:CT_alt}, $\delta_{1}$ can be unbiasedly estimated by a weighted average of DIDs. For all $t\in\{2,...,T\}$, for all $(d,d')\in (\mathcal{D}_1)^2$, and for all $\bm{d}^{-1}\in \mathcal{D}_2\times ... \times \mathcal{D}_K$, let
$$\mathcal{G}_{d,d',\bm{d}^{-1},t}=\left\{g:\ D^1_{g,t}=d,D^1_{g,t-1}=d',\bm{D}^{-1}_{g,t}=\bm{D}^{-1}_{g,t-1}=\bm{d}^{-1}\right\}$$
be the set of groups whose first treatment goes from $d'$ to $d$ from $t-1$ to $t$ while their other treatments are equal to $\bm{d}^{-1}$ at both dates. We then let $N_{d,d',\bm{d}^{-1},t}=\sum_{g\in \mathcal{G}_{d,d',\bm{d}^{-1},t}} N_{g,t}$ denote the total population of groups in $\mathcal{G}_{d,d',\bm{d}^{-1},t}$.
Let also
\begin{align}
\DID^f_{+,\bm{d}^{-1},t}&=\sum_{g\in \mathcal{G}_{1,0,\bm{d}^{-1},t}} \frac{N_{g,t}}{N_{1,0,\bm{d}^{-1},t}}\left(Y_{g,t}-Y_{g,t-1}\right) -\sum_{g\in \mathcal{G}_{0,0,\bm{d}^{-1},t}} \frac{N_{g,t}}{N_{0,0,\bm{d}^{-1},t}}\left(Y_{g,t}-Y_{g,t-1}\right),\label{eq:did+}\\
\DID^f_{-,\bm{d}^{-1},t}&=\sum_{g\in \mathcal{G}_{1,1,\bm{d}^{-1},t}}\frac{N_{g,t}}{N_{1,1,\bm{d}^{-1},t}}\left(Y_{g,t}-Y_{g,t-1}\right)- \sum_{g\in \mathcal{G}_{0,1,\bm{d}^{-1},t}}\frac{N_{g,t}}{N_{0,1,\bm{d}^{-1},t}}\left(Y_{g,t}-Y_{g,t-1}\right).\label{eq:did-}
\end{align}
Note that $\DID^f_{+,\bm{d}^{-1},t}$ is not defined when $N_{1,0,\bm{d}^{-1},t}=0$ or $N_{0,0,\bm{d}^{-1},t}=0$. In such instances, we let $\DID^f_{+,\bm{d}^{-1},t}=0$.  Similarly, we let $\DID^f_{-,\bm{d}^{-1},t}=0$ when $N_{1,1,\bm{d}^{-1},t}=0$ or $N_{0,1,\bm{d}^{-1},t}=0$.

\medskip
$\DID^f_{+,\bm{d}^{-1},t}$ compares the $t-1$-to-$t$ outcome evolution of groups whose first treatment goes from $0$ to $1$ from $t-1$ to $t$ while their other treatments are equal to $\bm{d}^{-1}$ at both dates, to the outcome evolution of groups whose first and other treatments are respectively equal to $0$ and $\bm{d}^{-1}$ at both dates. Under Assumption \ref{hyp:CT_alt}, the latter evolution is a valid counterfactual of the outcome evolution that the first groups would have experienced if their first treatment had remained equal to $0$ at period $t$. $\DID^f_{-,\bm{d}^{-1},t}$'s interpretation is similar, except that it compares groups whose first treatment is equal to $1$ at both dates to groups whose first treatment goes from $1$ to $0$.

\medskip
Finally, let
\begin{equation}
\DIDM^f = \sum_{t=2}^{T}\sum_{\bm{d}^{-1}\in \{0,1\}^{K-1}}\left(\frac{N_{1,0,\bm{d}^{-1},t}}{N_{\mathcal{S}_1}}\DID^f_{+,\bm{d}^{-1},t}+\frac{N_{0,1,\bm{d}^{-1},t}}{N_{\mathcal{S}_1}}\DID^f_{-,\bm{d}^{-1},t}\right)
	\label{eq:DIDM}
\end{equation}
if $N_{\mathcal{S}_1}>0$, and $\DIDM^f =0$ if $N_{\mathcal{S}_1}=0$. $\DIDM^f$ is just a weighted average of the $\DID^f_{+,\bm{d}^{-1},t}$ and $\DID^f_{-,\bm{d}^{-1},t}$ estimators, across values of the other treatments $\bm{d}^{-1}$ and across time periods $t$.
\begin{thm}\label{thm:alternative}
If Assumptions \ref{hyp:supp_gt}-\ref{hyp:independent_groups} and \ref{hyp:CT_alt} hold, $E\left[\DIDM^f\right]=\delta_{1}$.
\end{thm}

$\DIDM^f$ extends the $\DIDM$ estimator in \cite{dechaisemartin2020two} to settings with several treatments. With several treatments, one could show the analogue of Theorem \ref{thm:OVB} for the $\DIDM$ estimator in \cite{dechaisemartin2020two}: the fact that this estimator does not control for the other treatments may lead to a bias. To avoid that, the $\DIDM^f$ and $\DIDM$ estimators differ on three important dimensions: $\DIDM^f$ does not estimate the effect of the first treatment in $(g,t)$ cells such that at least one of $g$'s other treatments changes between $t-1$ and $t$; it drops control groups whose first treatment does not change but such that at least one of their other treatments changes between $t-1$ and $t$; and it compares switchers and non-switchers with the same baseline values of their other treatments. All those modifications ensure that our new estimator is not biased in the presence of other treatments with potentially heterogeneous treatment effects, but they may also come at a
cost in terms of precision: the $\DIDM^f$ estimator in this paper discards
several cells from the estimation. Accordingly, there may be a bias-variance trade-off between the two estimators.

\medskip
Like in \cite{dechaisemartin2020two}, it is straightforward to propose a placebo version of the $\DIDM^f$ estimator that one can use to test Assumption \ref{hyp:CT_alt}. To do so, one just needs to replace $Y_{g,t}-Y_{g,t-1}$ by $Y_{g,t-1}-Y_{g,t-2}$ in Equations \eqref{eq:did+} and \eqref{eq:did-} above, and exclude from the estimation groups experiencing a change in any of their treatments from $t-2$ to $t-1$. The resulting placebo estimator compares the outcome evolution of switchers and non-switchers, before switchers switch.

\medskip
The $\DIDM^f$ estimator can be extended to accommodate discrete non-binary treatments taking values in $\mathcal{D}_1= \{0,...,\overline{d}\}$, like the $\DIDM$ estimator in \cite{dechaisemartin2020two} \citep[see Web Appendix Section 4 of][]{dechaisemartin2020two}.
For all $t\in\{2,...,T\}$, for all $(d,d')\in (\mathcal{D}_1)^2$, and for all $\bm{d}^{-1}\in \mathcal{D}_2\times ... \times \mathcal{D}_K$, let
\begin{align*}
\DID^f_{d,d',\bm{d}^{-1},t} =\left[1\{d'<d\}-1\{d<d'\}\right] \bigg[ & \sum_{g\in \mathcal{G}_{d,d',\bm{d}^{-1},t}} \frac{N_{g,t}}{N_{d,d',\bm{d}^{-1},t}} [Y_{g,t}-Y_{g,t-1}] \\
& - \sum_{g\in \mathcal{G}_{d',d',\bm{d}^{-1},t}} \frac{N_{g,t}}{N_{d',d',\bm{d}^{-1},t}} [Y_{g,t}-Y_{g,t-1}]\bigg]
\end{align*}
be a DID estimator comparing the $t-1$-to-$t$ outcome evolution in groups whose first treatment changes from $d'$ to $d$ and whose other treatments are equal to $\bm{d}^{-1}$ at both dates, to the same outcome evolution in groups whose treatments do not change and with the same treatments in $t-1$. With a non-binary treatment, the $\DIDM^f$ estimator is a weighted average of the $\DID^f_{d,d',\bm{d}^{-1},t}$ estimators, across $d$, $d'$, $\bm{d}^{-1}$, and $t$, normalized by the average change of the first treatment among switchers, to ensure the estimator can be interpreted as an effect produced by a one-unit increase of the first treatment.

\medskip
Similarly, under Assumption \ref{hyp:CT_alt2}, and getting back to the binary treatment case, $\delta_{2}$ can be unbiasedly estimated by a weighted average of DIDs.
For all $t\in\{1,...,T-1\}$, for all $(d,d')\in (\mathcal{D}_1)^2$, and for all $\bm{d}^{-1}\in \mathcal{D}_2\times ... \times \mathcal{D}_K$,
let $N_{d,d',\bm{d}^{-1},t+1,t}=\sum_{g\in \mathcal{G}_{d,d',\bm{d}^{-1},t+1}} N_{g,t}$ denote the total population, at period $t$, of groups in $\mathcal{G}_{d,d',\bm{d}^{-1},t+1}$. Then, let
\begin{align*}
\DID^b_{+,\bm{d}^{-1},t}&=\sum_{g\in \mathcal{G}_{0,1,\bm{d}^{-1},t+1}} \frac{N_{g,t}}{N_{0,1,\bm{d}^{-1},t+1,t}}\left(Y_{g,t}-Y_{g,t+1}\right) -\sum_{g\in \mathcal{G}_{0,0,\bm{d}^{-1},t+1}} \frac{N_{g,t}}{N_{0,0,\bm{d}^{-1},t+1,t}}\left(Y_{g,t}-Y_{g,t+1}\right),\\
\DID^b_{-,\bm{d}^{-1},t}&=\sum_{g\in \mathcal{G}_{1,1,\bm{d}^{-1},t+1}}\frac{N_{g,t}}{N_{1,1,\bm{d}^{-1},t+1,t}}\left(Y_{g,t}-Y_{g,t+1}\right)- \sum_{g\in \mathcal{G}_{1,0,\bm{d}^{-1},t+1}}\frac{N_{g,t}}{N_{1,0,\bm{d}^{-1},t+1,t}}\left(Y_{g,t}-Y_{g,t+1}\right).
\end{align*}
In contrast to $\DID^f_{+,\bm{d}^{-1},t}$, which is a ``forward'' DID, $\DID^b_{+,\bm{d}^{-1},t}$ is a ``backward'' DID, from the future to the past. It compares the $t+1$-to-$t$ outcome evolution of groups whose first treatment goes from $0$ to $1$ from $t+1$ to $t$ while their other treatments are equal to $\bm{d}^{-1}$ at both dates, to the outcome evolution of groups whose first and other treatments are respectively equal to $0$ and $\bm{d}^{-1}$ at both dates. $\DID^b_{-,\bm{d}^{-1},t}$ has a similar interpretation, except that it compares groups whose first treatment is equal to $1$ at both dates to groups whose first treatment goes from $1$ to $0$ from $t+1$ to $t$.
Let
\begin{equation}
\DIDM^b = \sum_{t=1}^{T-1}\sum_{\bm{d}^{-1}\in \{0,1\}^{K-1}}\left(\frac{N_{0,1,\bm{d}^{-1},t+1,t}}{N_{\mathcal{S}_2}}\DID^b_{+,\bm{d}^{-1},t}+\frac{N_{1,0,\bm{d}^{-1},t+1,t}}{N_{\mathcal{S}_2}}\DID^b_{-,\bm{d}^{-1},t}\right)
	\label{eq:DIDMb}
\end{equation}
if $N_{\mathcal{S}_2}>0$, and $\DIDM^b =0$ if $N_{\mathcal{S}_2}=0$. By the exact same reaoning as in the proof of Theorem \ref{thm:alternative}, we obtain, under Assumptions \ref{hyp:supp_gt}-\ref{hyp:independent_groups} and \ref{hyp:CT_alt2}, $E\left[\DIDM^b\right]=\delta_{2}$.

\subsection{Additional results} 
\label{sub:additional_results}

In Web Appendix Section \ref{web:dyn}, we show that with a single treatment, $\DIDM^f$ can be used to estimate the effect of the contemporaneous value of the treatment, controlling for some lags of that treatment. Similarly, $\DIDM^b$ can be used to estimate the effect of a lag of the treatment, controlling for more recent lags and the treatment's contemporaneous value. We also show how to estimate dynamic effects with several binary and staggered treatments.

\medskip
In Web Appendix Section \ref{web:inf}, we consider both asymptotic and finite-sample inference. The asymptotic results are established under similar assumptions and arguments as those used to show the asymptotic normality of the $\DIDM$ estimator in \cite{dechaisemartin2020two} (see Theorem S6 in the Web Appendix therein), without any important conceptual difference. One limitation is that the asymptotic approximation may not be accurate. $\DIDM^f$ compares carefully selected treatment and control groups, and it could be the case that only a small number of groups can be included in those comparisons. We  deal with this issue by proposing confidence intervals that are exact in a finite sample of groups under a normality assumption, in the spirit of \cite{donald2007inference}. Though the exactness of those confidence intervals relies on strong conditions, we show that they remain asymptotically valid under much weaker assumptions.


\section{Application}\label{sec:appli}

In this section, we revisit \cite{hotz2011impact}.\footnote{This paper is the only one, in the census of TWFE papers published by the AER from 2010 to 2012 that we conducted in \cite{dechaisemartin2020two}, that has several treatments in the regression, relies at least partially on non-proprietary data, and for which the treatments are not continuous (thus making it possible to compute the $\DIDM^f$ estimator).} Unfortunately, many tables in this paper rely on proprietary data. The only table with TWFE regressions with several treatments that we can replicate is Table 11. Therefore, we focus on this table in our replication, though it is not the paper's main table.

\medskip
\cite{hotz2011impact} use a panel of the 50 US states and the District of Columbia, in 1987, 1992, and 1997, to estimate the effect of state center-based daycare regulations, namely the minimum years of schooling required to be the director of a center-based care and the minimum staff-to-child ratio, on the demand for family home daycare. Family home day cares are not subject to those regulations. More stringent regulations may increase the cost of center-based establishments, but may also increase their safety and quality. Accordingly, the effects of those regulations on the demand for family home daycare is ambiguous. The distributions of these regulations are shown in Table \ref{table:treatment_distrib}. The minimum years of schooling is a discrete treatment taking six values included between $0$ (no minimum) and $16$, with $14$ (associate degree) being the most frequent value. The minimum staff-to-child ratio is a also discrete treatment variable, taking seven values included between $0$ (no minimum) and 1/3 
(one professional per three children), with 1/4 
being the most frequent value.

\begin{table}[H]
\begin{center}
\caption{Distribution of the two treatments in \cite{hotz2011impact}}

\begin{tabular}{lc|lc}
\toprule
Min. years of schooling & \# of (g,t) cells & Min. staff-to-child ratio & \# of (g,t) cells \\ \midrule
0	 & 26 & 0 & 5\\
12	 & 36 & 1/8 & 2 \\
12.5 & 5 & 1/7 & 4 \\
13	 & 4 & 1/6 & 30 \\
14	 & 61 & 1/5 & 21 \\
16	 & 21 & 1/4 & 82 \\
 & & 1/3 & 9 \\
 \bottomrule
\end{tabular}
\label{table:treatment_distrib}
\end{center}
\end{table}

\cite{hotz2011impact} regress the revenue of family home day cares in state $g$ and year $t$ on state fixed effects, year fixed effects, 12 control variables, the minimum years of schooling required to be the director of a center-based care, the minimum staff-to-child ratio, and two indicators for whether there is no such minima, to allow for potentially non-linear effects. In Column (3) of their Table 11,
the coefficient on the minimum years of schooling treatment, $\widehat{\beta}^X_{fe}$, is equal to $-0.445$ and is highly significant (95\% confidence interval=$[-0.735,-0.155]$),\footnote{This confidence interval is slightly larger than that in \cite{hotz2011impact}, because we cluster standard errors at the state rather than at the state$\times$year level, which is more in line with the standard practice in empirical work \citep[see][]{bertrand2004}.} thus suggesting that increasing by one the years of schooling required for directors of center-based daycare decreases the revenue of family home daycare by 0.44 million USD.

\medskip
Dropping the 12 control variables from the regression does not affect that conclusion very much: the coefficient on the minimum years of schooling treatment, $\widehat{\beta}_{fe}$, is now equal to $-0.566$ and is still highly significant (95\% confidence interval=$[-0.852,-0.280]$). Below, we study $\widehat{\beta}_{fe}$, rather than $\widehat{\beta}^X_{fe}$, the coefficient estimated by \cite{hotz2011impact}. This is to ensure that the TWFE estimator we study is comparable to the $\DIDM^f$ estimator we compute below: while the $\DIDM^f$ estimator can be extended to allow for control variables, the sample on which it is computed in this application is not large enough to include 12 control variables.

\medskip
We now show that $\widehat{\beta}_{fe}$ may not be robust to heterogeneous effects across state and years, and may also be contaminated by the effects of the other treatments in the regression. Following Corollary \ref{cor:additive}, this coefficient  can be decomposed into the sum of four terms. The first term is a weighted sum of the effects of increasing by one the years of schooling required in 127 state$\times$year cells, where 44 effects receive a positive weight and 83 receive a negative weight, and where the positive and negative weights respectively sum to 7.897 and -6.897. The second term is a sum of the effects of not having a requirement on directors' years of schooling in 26 state$\times$year cells, where 11 effects receive a positive weight and 15 receive a negative weight, and where the positive and negative weights respectively sum to 0.148 and -0.148. The third term is a sum of the effects of increasing by one the staff to child ratio in 148 state$\times$year cells, where 51 effects  receive a positive weight and 97 receive a negative weight, and where the positive and negative weights respectively sum to 0.160 and -0.160. The last term is a sum of the effects of not having a requirement on staff to child ratio in 5 state$\times$year cells, where 4 effects receive a positive weight and 1 receive a negative weight, and where the positive and negative weights respectively sum to 0.055 and -0.055. Results are similar for the other three treatment coefficients in the regression, except that the contamination weights attached to them are even larger. For instance, for the coefficient on the staff to child ratio treatment, the weighted sum of the effects of the minimum years of schooling treatment has positive and negative weights summing to 246.222 and -246.222.

\medskip
When the other three treatment variables are dropped from the regression, the coefficient on the minimum years of schooling becomes small ($-0.020$) and insignificant (95\% confidence interval=$[-0.114,0.074]$). We follow Theorem 3 to decompose the coefficient in this ``short'' regression, and compare it to the coefficient in the ``long'' regression with the four treatments. The short regression's coefficient can be decomposed into the sum of four terms. The first term is  a weighted sum of the effects of increasing by one the years of schooling required in 127 state$\times$year cells, where 56 cells receive a positive weight and 71 receive a negative weight, and where the positive and negative weights respectively sum to 1.759 and -0.759. Thus, the short regression has considerably smaller negative weights in this first term than the long regression. The second term is a sum of the effects of not having a requirement on directors' years of schooling in 26 state$\times$year cells, where 5 effects receive a positive weight and 21 receive a negative weight, and where the positive and negative weights respectively sum to 0.008 and -0.077. The third term is a sum of the effects of increasing by one the staff to child ratio in 148 state$\times$year cells, where 61 effects  receive a positive weight and 87 receive a negative weight, and where the positive and negative weights respectively sum to 0.030 and -0.022. The last term is a sum of the effects of not having a requirement on staff to child ratio in 5 state$\times$year cells, where all effects receive a negative weight, and where the negative weights sum to -0.035. Thus, the short regression also has considerably less contamination weights than the long regression. Accordingly, the estimated maximal bias in Corollary \ref{cor:maxbiasshortlong} is almost five times lower for the short than for the long regression ($4.233 \times B$ versus $20.741\times B$), so the short regression is preferable per this maximal-bias metric.

\medskip
Finally, we compute the estimator proposed in Section \ref{sec:alternative_estimand}, for the minimum years of schooling treatment, controlling for the staff-to-child ratio treatment. Our estimators do not assume linear treatment effects, so we do not need to control for the indicators for whether there is no such minima.

\medskip
There are $127$ $(g,t)$ cells with a non-zero minimum years of schooling. On the other hand, there are only five $(g,t)$ cells in $\mathcal{S}_1$, all of which have  a non-zero minimum years of schooling. The five $(g,t)$ cells our estimator applies to are (Kentucky,1992), (Minnesota,1992), (Utah,1992), (Vermont,1992), and (Rhode Island,1997).\footnote{For the staff-to-child ratio treatment, the set $\mathcal{S}_1$ is even smaller as it only contains two $(g,t)$ cells. This is why we focus on the minimum-years-of-schooling treatment.} Of the 122 $(g,t)$ cells we lose when focusing on $\mathcal{S}_1$, 93 belong to states that do not experience any change of their minimum years of schooling, so their treatment effect cannot be identified under a parallel trends assumption. 24 $(g,t)$ cells  either also experience a change of their minimum staff-to-child ratio when their minimum years of schooling changes, or cannot be matched to a control state with the same baseline treatments. Then, estimating their treatment effect would require making constant effects assumptions. Finally, estimating the treatment of the remaining 5 $(g,t)$ cells would require assuming parallel trends over a longer horizon than over consecutive time periods.

\medskip
We find that $\DIDM^f=-0.029$. $\DIDM^f$ uses data from $5$ switching and $19$ control $(g,t)$ cells, so the asymptotic approximation in Web Appendix Section \ref{sub:asymptotic_approach} may not be very reliable for that estimator. Instead, we compute the exact confidence interval developed in Web Appendix Section \ref{sub:inference} and find that it is equal to $[-0.821, 0.807]$.\footnote{This confidence interval relies on Assumption \ref{hyp:non_overlap} in the Web Appendix, which does not hold in our data: Rhode Island and Washington are used twice in $\DIDM^f$. Removing these two states in one of the two $s$ they belong to (using the notation in Web Appendix Section \ref{sub:inference}) changes very slightly the value of $\DIDM^f$ (-0.0072 in lieu of -0.029).} In this application, the assumption that the first-differenced outcome is normally distributed is not rejected. We conduct a Shapiro-Wilk test separately for the 1987 to 1992 and for the 1992 to 1997 first differences, as the test assumes independent observations. None of the two tests is rejected (p-value= 0.98 and 0.46, respectively).

\medskip
To gain precision, one may further impose Assumption \ref{hyp:CT_alt2}. Doing so allows us to use $\DIDM^b$ to estimate the treatment effect in five $(g,t)$ cells in $\mathcal{S}_2$. $\mathcal{S}_1$ and $\mathcal{S}_2$ do not overlap and have the same numbers of cells, so we can also use $1/2(\DIDM^f+\DIDM^b)$ to estimate $\delta$, the average treatment effect in $\mathcal{S}_1\cup \mathcal{S}_2$. We find that $1/2(\DIDM^f+\DIDM^b)=-0.016$. $1/2(\DIDM^f+\DIDM^b)$ uses data from $50$ $(g,t)$ cells, coming from $30$ different states. The asymptotic approximation in Web Appendix Section \ref{sub:asymptotic_approach} may be more reasonable for that estimator,\footnote{To verify that, we considered simulations with the same design as in the application but with no effects of the treatments, and $(\Delta Y_{g,2}(\0), \Delta Y_{g,3}(\0))$ drawn either from a normal distribution $\mathcal{N}(\0,\Sigma)$, with $\Sigma$ equal to the estimated variance matrix on the sample, or from  the empirical distribution of $(\Delta Y_{g,2}, \Delta Y_{g,3})$. In both cases, the coverage of our confidence interval was higher than 95\% (95.4\% and 99.3\%, respectively).}  so we follow Theorem \ref{thm:asym} therein to compute a 95\% confidence interval for $\delta$. We find that this confidence interval is equal to $[-0.126, 0.094]$. We also test the equality between $\delta$ and $\beta_{fe}$, and reject the null hypothesis at all conventional levels (p-value=$4\times 10^{-4}$). Hence, as discussed above, we can reject the hypothesis that the effects of the minimum years of schooling and staff-to-child ratio treatments are homogenous.

\begin{table}[H]
\begin{center}
\caption{Estimators of the effect of the minimum years of schooling treatment}
\begin{tabular}{l c c c c }
\toprule
 & Estimate & 95\% Confidence Interval \\ \midrule
$\widehat{\beta}^X_{fe}$ 	 & $-0.445$ & $[-0.735,-0.155]$ \\
$\widehat{\beta}_{fe}$	 & $-0.566$ & $[-0.860,-0.272]$  \\
$\widehat{\beta}_{s}$	 & $-0.022$ & $[-0.117,0.077]$  \\
$\DIDM^f$ & $-0.029$ &  $[-0.821, 0.807]$  \\
$1/2(\DIDM^f+\DIDM^b)$ & $-0.016$ & $[-0.126, 0.094]$  \\
\bottomrule
\end{tabular}\label{table:results}
\end{center}
\end{table}

**  [95

Let us summarize our results. Using a TWFE regression with several treatments, \cite{hotz2011impact} find that increasing the years of schooling required for directors of center-based daycare significantly decreases the revenue of family home daycare. We show that in the presence of heterogeneous treatment effects, their regression estimates a highly-non-convex combination of the effects of the years of schooling treatment, and is contaminated by the effects of the other treatments. Therefore, their finding may not be robust to heterogeneous treatment effects. Then, we use our robust estimators to assess if, in the presence of heterogeneous effects, one can conclude, for at least a subset of $(g,t)$ cells, that increasing the years of schooling requirement significantly decreases the revenue of family home daycare. The answer is negative, as our estimators are insignificant. Moreover, one of our estimators is significantly different from the TWFE estimator, thus allowing us to reject the null hypothesis that the effects of all treatments are constant in this application. Overall, there is no evidence that the finding in \cite{hotz2011impact} is robust to heterogeneous effects, while there is evidence that treatment effects are heterogeneous in this application.

\section{Conclusion}

In this paper, we show that treatment coefficients in TWFE regressions with several treatments may not be robust to heterogeneous effects, and could be contaminated by the effects of other treatments in the regression. We propose alternative DID estimators that are robust to heterogeneous effects and do not suffer from this contamination problem.

\newpage
\bibliography{biblio}

\newpage
\appendix

\section{Proofs}

\subsection{Theorem \ref{thm:main}} 
\label{sub:theorem_ref_thm_main}

The result directly follows from Theorem \ref{thm:extension}. If $K=2$, $\bm{D}^{-1}_{g,t}=D^2_{g,t}$. Then, $\bm{D}^{-1}_{g,t}\neq \bm{0}^{-1}$ if and only if $D^2_{g,t}=1$, and one then has $D^2_{g,t}\Delta^{-1}_{g,t}=D^2_{g,t}\Delta^{2}_{g,t}$.

\subsection{Theorem \ref{thm:extension}} 
\label{sub:theorem_ref_thm_extension}

We first establish the following lemma.
\begin{lem}\label{lem:extension}
If Assumptions \ref{hyp:supp_gt}-\ref{hyp:common_trends} hold, for all $(g,g',t,t')\in \{1,...,G\}^2\times \{1,...,T\}^2,$
\begin{align*}
&E\left(Y_{g,t}\middle|\bm{D}\right)-E\left(Y_{g,t'}\middle|\bm{D}\right)-\left(E\left(Y_{g',t}\middle|\bm{D}\right)-E\left(Y_{g',t'}\middle|\bm{D}\right)\right)\\
=&D^1_{g,t}E\left(\Delta^{1}_{g,t}\middle|\bm{D}\right) + E\left(\Delta^{-1}_{g,t}\middle|\bm{D}\right) -  D^1_{g',t}E\left(\Delta^1_{g',t}(\bm{D}^{-1}_{g',t})\middle|\bm{D}\right) - E\left(\Delta^{-1}_{g',t}\middle|\bm{D}\right)  \\
-& D^1_{g,t'}E\left(\Delta^1_{g,t'}(\bm{D}^{-1}_{g,t'})\middle|\bm{D}\right) - E\left(\Delta^{-1}_{g,t'}\middle|\bm{D}\right)
+ D^1_{g',t'}E\left(\Delta^1_{g',t'}(\bm{D}^{-1}_{g',t'})\middle|\bm{D}\right) + E\left(\Delta^{-1}_{g',t'}\middle|\bm{D}\right).
\end{align*}
\end{lem}

\subsubsection*{Proof of Lemma \ref{lem:extension}}

For all $(g,t)\in \{1,...,G\}\times \{1,...,T\},$
\begin{align}
E\left(Y_{g,t}\middle|\bm{D}\right)
=&E\bigg(Y_{g,t}(0,\bm{0}^{-1})+D^1_{g,t} (Y_{g,t}(1,\bm{D}^{-1}_{g,t})-Y_{g,t}(0,\bm{D}^{-1}_{g,t})+ Y_{g,t}(0,\bm{D}^{-1}_{g,t}) - Y_{g,t}(0,\bm{0}^{-1}))\notag \\
&+(1-D^1_{g,t})(Y_{g,t}(0,\bm{D}^{-1}_{g,t}) -Y_{g,t}(0,\bm{0}^{-1}))\bigg| \bm{D}\bigg) \notag \\
=&E\left(Y_{g,t}(0,\bm{0}^{-1})\middle|\bm{D}\right)+D^1_{g,t}E\left(\Delta^{1}_{g,t} \middle|\bm{D}\right)+ E\left(\Delta^{-1}_{g,t} \middle|\bm{D}\right)\notag \\
=&E\left(Y_{g,t}(0,\bm{0}^{-1})\middle|\bm{D}_g\right)+D^1_{g,t}E\left(\Delta^{1}_{g,t} \middle|\bm{D}\right)+ E\left(\Delta^{-1}_{g,t} \middle|\bm{D}\right),
\label{eq:1ststep}
\end{align}
where the last equality follows from Assumption \ref{hyp:independent_groups}. Moreover, by Assumption \ref{hyp:common_trends}
\begin{align}
	& E\left(Y_{g,t}(0,\bm{0}^{-1})\middle|\bm{D}_g\right) - E\left(Y_{g,t'}(0,\bm{0}^{-1})\middle|\bm{D}_g\right) - E\left(Y_{g',t}(0,\bm{0}^{-1})\middle|\bm{D}_g\right) + E\left(Y_{g',t'}(0,\bm{0}^{-1})\middle|\bm{D}_g\right) \notag \\
	= & 0. \label{eq:2ndstep}
\end{align}
The result follows by combining \eqref{eq:1ststep} and \eqref{eq:2ndstep}.

\subsubsection*{Proof of Theorem \ref{thm:extension}}

It follows from the Frisch-Waugh theorem and the definition of $\eps_{g,t}$ that
\begin{equation}\label{eq:eq1_beta1_sharp}
E\left(\widehat{\beta}_{fe}\middle|\bm{D}\right)=\frac{\sum_{g,t}N_{g,t}\eps_{g,t}E\left(Y_{g,t}\middle|\bm{D}\right)}{\sum_{g,t}N_{g,t}\eps_{g,t}D^1_{g,t}}.
\end{equation}
Now, by definition of $\eps_{g,t}$ again,
\begin{align}
&\sum_{t=1}^{T}N_{g,t}\eps_{g,t}=0\text{ for all }g\in \{1,...,G\}, \label{eq:residu_centreG}\\
&\sum_{g=1}^{G}N_{g,t}\eps_{g,t}=0\text{ for all }t\in \{1,...,T\}, \label{eq:residu_centreT}.
\end{align}
Then,
\begin{align}
& \sum_{g,t}N_{g,t}\eps_{g,t}E\left(Y_{g,t}\middle|\bm{D}\right) \nonumber \\
=&  \sum_{g,t}N_{g,t}\eps_{g,t}\left(E\left(Y_{g,t}\middle|\bm{D}\right)-E\left(Y_{g,1}\middle|\bm{D}\right)-E\left(Y_{1,t}\middle|\bm{D}\right)+E\left(Y_{1,1}\middle|\bm{D}\right)\right) \nonumber\\
=&  \sum_{g,t}N_{g,t}\eps_{g,t}\left(D^1_{g,t}E\left(\Delta^{1}_{g,t}\middle|\bm{D}\right) + E\left(\Delta^{-1}_{g,t}\middle|\bm{D}\right) -  D^1_{1,t}E\left(\Delta^1_{1,t}(\bm{D}^{-1}_{1,t})\middle|\bm{D}\right) - E\left(\Delta^{-1}_{1,t})\middle|\bm{D}\right)  \right.\nonumber \\
-& \left. D^1_{g,1}E\left(\Delta^1_{g,1}(\bm{D}^{-1}_{g,1})\middle|\bm{D}\right) - E\left(\Delta^{-1}_{g,1})\middle|\bm{D}\right) +  D^1_{1,1}E\left(\Delta^1_{1,1}(\bm{D}^{-1}_{1,1})\middle|\bm{D}\right) + E\left(\Delta^{-1}_{1,1})\middle|\bm{D}\right) \right)\nonumber \\
=&  \sum_{g,t}N_{g,t}\eps_{g,t}\left(D^1_{g,t}E\left(\Delta^{1}_{g,t}\middle|\bm{D}\right) + E\left(\Delta^{-1}_{g,t}\middle|\bm{D}\right) \right) \nonumber \\
=&  \sum_{(g,t):D^1_{g,t}=1}N_{g,t}\eps_{g,t}E\left(\Delta^{1}_{g,t}\middle|\bm{D}\right)+\sum_{(g,t):\bm{D}^{-1}_{g,t}\ne \bm{0}^{-1}}N_{g,t}\eps_{g,t}E\left(\Delta^{-1}_{g,t} \middle|\bm{D}\right).
\label{eq:num_beta1_sharp}
\end{align}
The first and third equalities follow from Equations \eqref{eq:residu_centreG} and \eqref{eq:residu_centreT}.
The second equality follows from Lemma \ref{lem:extension}. The fourth equality follows from the fact that $\Delta^0_{g,t}(\bm{0}^{-1})=0$. Finally,
\begin{equation}
\sum_{g,t}N_{g,t}\eps_{g,t}D^1_{g,t}=\sum_{(g,t):D^1_{g,t}=1}N_{g,t}\eps_{g,t}.	
	\label{eq:denom_beta1_sharp}
\end{equation}
Combining \eqref{eq:eq1_beta1_sharp}, \eqref{eq:num_beta1_sharp}, \eqref{eq:denom_beta1_sharp} yields
\begin{equation}\label{eq:thm1_cond}
E\left(\widehat{\beta}_{fe}\middle|\mathbf{D}\right) =  \sum_{(g,t):D^1_{g,t}=1}\frac{N_{g,t}}{N_1}w_{g,t} E\left(\Delta^{1}_{g,t}\middle|\bm{D}\right)+\sum_{(g,t):\bm{D}^{-1}_{g,t}\neq \bm{0}^{-1}}\frac{N_{g,t}}{N_1}w_{g,t} E\left(\Delta^{-1}_{g,t}\middle|\bm{D}\right).
\end{equation}
Then, the first result follows from the law of iterated expectations. Finally, if $K=2$ or the treatments are mutually exclusive,
$$\sum_{(g,t):\bm{D}^{-1}_{g,t}\ne \bm{0}^{-1}}N_{g,t}\eps_{g,t}E\left(\Delta^{-1}_{g,t} \middle|\bm{D}\right) =
\sum_{k=2}^K \sum_{(g,t):D^k_{g,t}=1}N_{g,t}\eps_{g,t}E\left(\Delta^{-1}_{g,t} \middle|\bm{D}\right).$$
Moreover, by definition of  $\eps_{g,t}$, $\sum_{(g,t):D^k_{g,t}=1}N_{g,t}\eps_{g,t}=0$ for all $k=2,...,K-1$. The second result follows.

\subsection{Theorem \ref{thm:OVB}}

The proof is the same as 
that of Theorem \ref{thm:main}, with just one difference: we do not have
$\sum_{(g,t):D^2_{g,t}=1} N_{g,t} $ $\times \eps^s_{g,t}=0$, since $\eps^s_{g,t}$ is not orthogonal to $D^2_{g,t}$ in general.

\subsection{Corollary \ref{cor:maxbiasshortlong}}

The result directly follows from Theorems \ref{thm:main} and \ref{thm:OVB}, the triangle inequality, and the fact there is a real number $B$ such that $|\Delta^{1}_{g,t}|\leq B$ and $|\Delta^{2}_{g,t}|\leq B$ for all $(g,t)$. The first bound is reached when $\Delta^{1}_{g,t}=B\times (2\times1\{w_{g,t}\geq 1\}-1)$ and $\Delta^{2}_{g,t}=B(2\times1\{w_{g,t}\geq 0\}-1)$, the second bound is reached when $\Delta^{1}_{g,t}=B\times (2\times1\{w^s_{g,t}\geq 1\}-1)$ and $\Delta^{2}_{g,t}=B(2\times1\{w^s_{g,t}\geq 0\}-1).$


\subsection{Proposition \ref{prop:maximal_set}} 
\label{proof:prop_max}

The set $\mathcal{T}_1\backslash\mathcal{S}_1$ can be partitioned into three groups, $\mathcal{V}_1,\mathcal{V}_2,\mathcal{V}_3$, defined by:
\begin{align*}
	\mathcal{V}_1 & = \left\{(g,t): D^1_{g,t}=1 \text{ and either } D^1_{g,t-1}=1 \text{ or } t=1\right\}, \\
	\mathcal{V}_2 & = \left\{(g,t): D^1_{g,t}=1,\, D^1_{g,t-1}=0 \text{ and } \bm{D}^{-1}_{g,t}\ne \bm{D}^{-1}_{g,t-1}\right\}, \\
	\mathcal{V}_3 & = \left\{(g,t): D^1_{g,t}=1,\,D^1_{g,t-1}=0, \, \bm{D}^{-1}_{g,t}=\bm{D}^{-1}_{g,t-1}, \, \forall g'\ne g, \text{ either } \bm{D}_{g',t} \ne \bm{D}_{g',t-1} \right.\\
	& \qquad \text{ or } \bm{D}_{g',t-1} \ne \bm{D}_{g,t-1}\big\}.
\end{align*}
Since the assumptions impose no joint restrictions on these three groups, it suffices to prove that $\delta_{\mathcal{V}} = E[\sum_{(g,t)\in \mathcal{V}} (N_{g,t}/N_{\mathcal{V}}) \Delta^{1}_{g,t}]$ is not identified if $\mathcal{V} \subset \mathcal{V}_k$ ($k=1,...,3$). To this end, fix $c\ne 0$ and for all $(g,t)\in\mathcal{V}$, let $\widetilde{Y}_{g,t}(0,\bm{D}^{-1}_{g,t})=Y_{g,t}(0,\bm{D}^{-1}_{g,t})+c$,  $\widetilde{Y}_{g,t}(\bm{d})=Y_{g,t}(\bm{d})$ for all $\bm{d}\ne (0,\bm{D}^{-1}_{g,t})$ and $\widetilde{Y}_{g',t'}(\bm{d})=Y_{g',t'}(\bm{d})$ for all $\bm{d}$ and $(g',t')\not\in\mathcal{V}$.

\medskip
If $\mathcal{V} \subset \mathcal{V}_1$, Assumption \ref{hyp:CT_alt} does not impose any restriction on $Y_{g,t}(0,\bm{D}^{-1}_{g,t})$ for $(g,t)\in\mathcal{V}_1$, since either $D^1_{g,t-1}=1$ or $t=1$. The potential outcomes $(\widetilde{Y}_{g,t}(\bm{d}))_{(g,t,\bm{d})}$ are thus compatible with the data  and Assumptions \ref{hyp:supp_gt}-\ref{hyp:independent_groups} and \ref{hyp:CT_alt}. Since they lead to $\widetilde{\delta}_{\mathcal{V}} = \delta_{\mathcal{V}}-c$, $\delta_{\mathcal{V}}$ is not identified.

\medskip
Now, if $\mathcal{V} \subset \mathcal{V}_2$, Assumption \ref{hyp:CT_alt} imposes a restriction on $Y_{g,t}(0,\bm{D}^{-1}_{g,t-1})-Y_{g,t-1}(0,\bm{D}^{-1}_{g,t-1})$ for $(g,t)\in\mathcal{V}$. But since $\bm{D}^{-1}_{g,t}\ne \bm{D}^{-1}_{g,t-1}$ for such cells, we have
$$\widetilde{Y}_{g,t}(0,\bm{D}^{-1}_{g,t-1})-\widetilde{Y}_{g,t-1}(0,\bm{D}^{-1}_{g,t-1})=Y_{g,t}(0,\bm{D}^{-1}_{g,t-1})-Y_{g,t-1}(0,\bm{D}^{-1}_{g,t-1}),$$
using here $(g,t-1)\not\in\mathcal{V}_2$. Therefore, $\widetilde{Y}_{g,t}(0,\bm{D}^{-1}_{g,t-1})- \widetilde{Y}_{g,t-1}(0,\bm{D}^{-1}_{g,t-1})$ satisfies the restriction of Assumption \ref{hyp:CT_alt}. So again, $(\widetilde{Y}_{g,t}(\bm{d}))_{(g,t,\bm{d})}$ are compatible with the data and Assumptions \ref{hyp:supp_gt}-\ref{hyp:independent_groups} and \ref{hyp:CT_alt}, and  $\delta_{\mathcal{V}}$  is not identified.

\medskip
Finally, assume $\mathcal{V} \subset \mathcal{V}_3$. In this case, $(\widetilde{Y}_{g,t}(\bm{d}))_{(g,t,\bm{d})}$ may violate Assumption \ref{hyp:CT_alt}: if $\bm{D}_{g',t-1}=\bm{D}_{g,t-1}=(0,\bm{d}^{-1})$ for some $(g,t)\in \mathcal{V}$ and $(g',t)\not\in\mathcal{V}_3$, we have
\begin{align*}
& E[\widetilde{Y}_{g',t}(0,\bm{d}^{-1}) - \widetilde{Y}_{g',t-1}(0,\bm{d}^{-1})|\bm{D}_{g',t-1}=(0,\bm{d}^{-1})] \\
= & E[Y_{g',t}(0,\bm{d}^{-1}) - Y_{g',t-1}(0,\bm{d}^{-1})|\bm{D}_{g',t-1}=(0,\bm{d}^{-1})] \\
= & E[Y_{g,t}(0,\bm{d}^{-1}) - Y_{g,t-1}(0,\bm{d}^{-1})|\bm{D}_{g,t-1}=(0,\bm{d}^{-1})] \\
= & E[\widetilde{Y}_{g,t}(0,\bm{d}^{-1}) - \widetilde{Y}_{g,t-1}(0,\bm{d}^{-1})|\bm{D}_{g,t-1}=(0,\bm{d}^{-1})] - c.	
\end{align*}
To fix this, we simply let $\widetilde{Y}_{g',t}(0,\bm{d}^{-1})=Y_{g',t}(0,\bm{d}^{-1})+c$ for such $(g',t)\not\in\mathcal{V}_3$. This change is still compatible with the data: by definition of $\mathcal{V}_3$ and because $\bm{D}_{g',t-1}=\bm{D}_{g,t-1}$, we have $\bm{D}_{g',t}\ne (0,\bm{d}^{-1})$. Hence, with this modification, $(\widetilde{Y}_{g,t}(\bm{d}))_{(g,t,\bm{d})}$ are compatible with the data and Assumptions \ref{hyp:supp_gt}-\ref{hyp:independent_groups} and \ref{hyp:CT_alt}. This shows that again, $\delta_{\mathcal{V}}$  is not identified.


\subsection{Theorem \ref{thm:alternative}} 
\label{sub:proof_of_theorem_ref_thm_alternative}

First, by definition of $\DIDM^f$,
\begin{align}\label{eq:DIDM2}
\DIDM^f=&\sum_{t=2}^{T}\sum_{\bm{d}^{-1}\in \{0,1\}^{K-1}}\frac{N_{1,0,\bm{d}^{-1},t}}{N_{\mathcal{S}_1}}\DID^f_{+,\bm{d}^{-1},t}+\frac{N_{0,1,\bm{d}^{-1},t}}{N_{\mathcal{S}_1}}\DID^f_{-,\bm{d}^{-1},t},
\end{align}
using here the convention that $0/0=0$. Let $t\geq 2$ and $\bm{d}^{-1}\in \{0,1\}^{K-1}$ be such that $N_{1,0,\bm{d}^{-1},t}>0$ and $N_{0,0,\bm{d}^{-1},t}>0$. For every $g$ such that $D^1_{g,t-1}=0$, $D^1_{g,t}=1$, and $\bm{D}^{-1}_{g,t}=\bm{D}^{-1}_{g,t-1}=\bm{d}^{-1}$, we have
\begin{align}
	E\left(Y_{g,t}-Y_{g,t-1}\middle|\bm{D}\right) = & E\left(\Delta^{1}_{g,t}\middle|\bm{D}\right) + E\left(Y_{g,t}(0,\bm{d}^{-1})-Y_{g,t-1}(0,\bm{d}^{-1})\middle|\bm{D}\right). \label{eq:for_TC1}
\end{align}
Under Assumptions \ref{hyp:independent_groups} and \ref{hyp:CT_alt}, for all $t\geq 2$, there exists $\psi_{0,\bm{d}^{-1},t}\in\R$ such that for all $g \in \mathcal{G}_{0, 0, \bm{d}^{-1}, t}\cup \mathcal{G}_{1, 0, \bm{d}^{-1}, t}$,
\begin{align}
E\left(Y_{g,t}(0,\bm{d}^{-1})-Y_{g,t-1}(0,\bm{d}^{-1})\middle|\bm{D}\right)
=&E\left(Y_{g,t}(0,\bm{d}^{-1})-Y_{g,t-1}(0,\bm{d}^{-1})\middle|\bm{D}_g\right) \nonumber \\
=& E\left(Y_{g,t}(0,\bm{d}^{-1})-Y_{g,t-1}(0,\bm{d}^{-1})\middle| D^1_{g,t-1}=0, \bm{D}^{-1}_{g,t-1}=\bm{d}^{-1} \right) \nonumber \\
=&\psi_{0,\bm{d}^{-1},t}.\label{eq:phi}
\end{align}
As a result,
\begin{align*}
	&N_{1,0,\bm{d}^{-1},t}E\left(\DID^f_{+,\bm{d}^{-1},t}\middle|\bm{D}\right)\\
=& \sum_{g\in \mathcal{G}_{1,0,\bm{d}^{-1},t}} N_{g,t}E\left(\Delta^{1}_{g,t}\middle|\bm{D}\right)
+ \sum_{g\in \mathcal{G}_{1,0,\bm{d}^{-1},t}} N_{g,t}E\left(Y_{g,t}(0,\bm{d}^{-1})-Y_{g,t-1}(0,\bm{d}^{-1})\middle|\bm{D}\right)  \\
	& - \frac{N_{1,0,\bm{d}^{-1},t}}{N_{0,0,\bm{d}^{-1},t}}\sum_{g\in \mathcal{G}_{0,0,\bm{d}^{-1},t}} N_{g,t} E\left(Y_{g,t}(0,\bm{d}^{-1})-Y_{g,t-1}(0,\bm{d}^{-1})\middle|\bm{D}\right) \\
	= & \sum_{g\in \mathcal{G}_{1,0,\bm{d}^{-1},t}} N_{g,t}E\left(\Delta^{1}_{g,t}\middle|\bm{D}\right)
+\psi_{0,\bm{d}^{-1},t}\left(\sum_{g\in \mathcal{G}_{1,0,\bm{d}^{-1},t}} N_{g,t}-\frac{N_{1,0,\bm{d}^{-1},t}}{N_{0,0,\bm{d}^{-1},t}}\sum_{g\in \mathcal{G}_{0,0,\bm{d}^{-1},t}} N_{g,t}\right)\\
	= & \sum_{g\in \mathcal{G}_{1,0,\bm{d}^{-1},t}} N_{g,t}E\left(\Delta^{1}_{g,t}\middle|\bm{D}\right).
\end{align*}
The first equality follows by \eqref{eq:for_TC1}, the second by \eqref{eq:phi}, and the third after some algebra. Given that $\DID^f_{+,\bm{d}^{-1},t}=0$ if $N_{1,0,\bm{d}^{-1},t}=0$ or $N_{0,0,\bm{d}^{-1},t}=0$, we obtain, by definition of $\mathcal{S}_1$ and with the convention that sums over empty sets are 0,
\begin{equation}
E\left(N_{1,0,\bm{d}^{-1},t}\DID^f_{+,\bm{d}^{-1},t}\middle|\bm{D}\right) = E\bigg(\sum_{\substack{g:D^1_{g,t}=1,\bm{D}^{-1}_{g,t}=\bm{d}^{-1} \\ (g,t)\in\mathcal{S}_1}} N_{g,t}\Delta^{1}_{g,t}\bigg|\bm{D}\bigg).	
	\label{eq:DIDplus}
\end{equation}


\medskip
A similar reasoning yields, for all $t\geq 2$ and $\bm{d}^{-1}\in \{0,1\}^{K-1}$,
\begin{equation}
E\left(N_{0,1,\bm{d}^{-1},t}\DID^f_{-,\bm{d}^{-1},t}\middle|\bm{D}\right) =E\bigg( \sum_{\substack{g:D^1_{g,t}=0,\bm{D}^{-1}_{g,t}=\bm{d}^{-1} \\ (g,t)\in\mathcal{S}_1}} N_{g,t}\Delta^{1}_{g,t}\bigg|\bm{D}\bigg).	
\label{eq:DIDminus}
\end{equation}
Plugging \eqref{eq:DIDplus} and \eqref{eq:DIDminus} into \eqref{eq:DIDM2} yields
\begin{align*}
	E(\DIDM^f) = & E\bigg(E\bigg(\sum_{t=2}^{T}\sum_{\bm{d}^{-1}\in \{0,1\}^{K-1}} \sum_{\substack{g:\bm{D}^{-1}_{g,t}=\bm{d}^{-1} \\ (g,t)\in\mathcal{S}_1}} N_{g,t}\Delta^{1}_{g,t}\bigg|\bm{D}\bigg)\bigg)\\
	= & E\bigg(E\bigg(\sum_{(g,t)\in\mathcal{S}_1} N_{g,t}\Delta^{1}_{g,t}\bigg|\bm{D}\bigg)\bigg)\\
	= & \delta_{1}.
\end{align*}



\newpage
\begin{center}
{\Large Web Appendix of ``Two-way Fixed Effects and Differences-in-Differences Estimators with Several Treatments''}	
\end{center}

\appendix

\section{Replacing Assumption \ref{hyp:common_trends} by Assumption \ref{hyp:CT_alt} may exacerbate the problems of TWFE regressions.}\label{appendix:decompo_altCT}

Under Assumption \ref{hyp:CT_alt}, we cannot in general write $\beta_{fe}$ as a function of the design and treatment effects only, and replacing Assumption \ref{hyp:common_trends} by Assumption \ref{hyp:CT_alt} may actually exacerbate the problems of TWFE regressions: in addition to not being robust to heterogeneous treatment effects, TWFE regressions may now be biased even with homogenous treatment effects. This point is not specific to the case with several treatments we consider in the paper, so let us momentarily assume that $K=1$. Then, assume that we have two groups and three periods, and that almost surely, the first group is treated at $t=3$, while the second group is treated at $t=2$ and $t=3$. This is the same example as in \cite{dechaisemartin2020two}. Let $\Delta_{g,t}=Y_{g,t}(1)-Y_{g,t}(0)$. Then
$$\widehat{\beta}_{fe} = \frac{1}{2} (\DID_1 + \DID_2),$$
with $\DID_1=Y_{2,2}-Y_{2,1} - (Y_{1,2}-Y_{1,1})$ and $\DID_2=Y_{1,3}-Y_{1,2} - (Y_{2,3}-Y_{2,2})$. Then, under Assumption \ref{hyp:CT_alt}, we still have that $E[\DID_1]=E[\Delta_{2,2}]$. But under Assumption \ref{hyp:CT_alt},
\begin{align*}
E[\DID_2]=&E[Y_{1,3}(1)-Y_{1,2}(0) - (Y_{2,3}(1)-Y_{2,2}(1))]\\
=&E[\Delta_{1,3}]+E[\Delta_{2,2}]-E[\Delta_{2,3}]\\
+&E[Y_{1,3}(0)]-E[Y_{1,2}(0)]-(E[Y_{2,3}(0)]-E[Y_{2,2}(0)]).
\end{align*}
Therefore,
\begin{align*}
E[\widehat{\beta}_{fe}]=&\frac{1}{2}E[\Delta_{1,3}]+E[\Delta_{2,2}]-\frac{1}{2}E[\Delta_{2,3}]\nonumber\\
+&\frac{1}{2}(E[Y_{1,3}(0)]-E[Y_{1,2}(0)]-(E[Y_{2,3}(0)]-E[Y_{2,2}(0)])).
\end{align*}
The first term is the decomposition of $\widehat{\beta}_{fe}$ as a weighted sum of treatment effects in Theorem 1 in \cite{dechaisemartin2020two}. Under Assumption \ref{hyp:common_trends}, the second term in equal to zero. However, under  Assumption \ref{hyp:CT_alt} that second term may not be equal to zero: because $D_{1,2}=0\ne D_{2,2}$,  Assumption \ref{hyp:CT_alt} does not impose anything on $E[Y_{1,3}(0)]-E[Y_{1,2}(0)]-(E[Y_{2,3}(0)]-E[Y_{2,2}(0)])$. Hence, replacing Assumption \ref{hyp:common_trends} by Assumption \ref{hyp:CT_alt} may actually exacerbate the problems of TWFE regressions. In addition to not being robust to heterogeneous treatment effects, TWFE regressions may now be biased even with homogenous treatment effects.
	
\section{Cells belonging to $\mathcal{T}_1$ and not to $\mathcal{S}_1$}\label{appendix:diffATTLATE}

The cells belonging to $\mathcal{T}_1$ and not to $\mathcal{S}_1$ satisfy one of the five mutually exclusive conditions below.

\medskip
\textbf{Condition 1:} $(g,t)$ is such that $D^1_{g,t}=1$ for all $t$. As this cell's first treatment never changes, its effect cannot be identified under a parallel trends assumption.

\medskip
\textbf{Condition 2:} $(g,t)$ is such that $D^1_{g,t}=1$, $D^1_{g,t-1}=1$ or $t=1$, and $D^1_{g,t+1}=0$. To estimate the first-treatment's effect in cell $(g,t)$, one could compare $g$'s $t+1$-to-$t$ outcome evolution to that of another group experiencing no treatment change and with the same treatments at period $t+1$. This estimator is unbiased for the first-treatment's effect in cell $(g,t)$ under Assumption \ref{hyp:CT_alt2}, but not under Assumption \ref{hyp:CT_alt}.

\medskip
\textbf{Condition 3:} $(g,t)$ is such that $D^1_{g,t}=1$, $D^1_{g,t-1}=1$ or $t=1$, and $D^1_{g,t'}=0$ for a $t'$ lower than $t-2$ or greater than $t+2$. To identify the effect of the first treatment in one such cell, one could compare $g$'s $t'$-to-$t$ outcome evolution to that of another group $g'$ with the same treatments at $t'$ and $t$ ($\bm{D}_{g',t}=\bm{D}_{g',t'}$) and with the same treatments as $g$ in $t'$ ($\bm{D}_{g,t'}=\bm{D}_{g',t'}$). However, such a comparison relies on a parallel trends assumption over a longer time horizon than Assumption \ref{hyp:CT_alt}, which only imposes parallel trends over consecutive periods. One may argue that if one is ready to impose parallel trends over consecutive periods, the cost of further imposing parallel trends over longer horizons is minimal. But if one views parallel trends as a reasonable first-order approximation, rather than an assumption that exactly holds, this first-order approximation may become poorer over longer horizons, for instance if there are group specific linear trends \citep[see][]{roth2019pre,rambachan2019honest}. Accordingly, estimators relying on long-run parallel trends assumptions may be more biased than estimators relying on parallel trends over consecutive periods.

\medskip
\textbf{Condition 4:} $(g,t)$ is such that $D^1_{g,t}=1$, $D^1_{g,t-1}=0$, and $\bm{D}^{-1}_{g,t}\ne \bm{D}^{-1}_{g,t-1}$, meaning that one of $g$'s other treatments also changes from $t-1$ to $t$. To identify the effect of the first treatment in one such cell, one could compare $g$'s outcome evolution to that of another group experiencing no change of its first treatment and the same changes of its other treatments as $g$, like in the example in Equation \eqref{eq:example2DID2}. However, such a comparison is only valid if the effect of the other treatments is constant between groups, as discussed in Section \ref{subsub:intuition}.

\medskip
\textbf{Condition 5:} $(g,t)$ is such that $D^1_{g,t}=1$, $D^1_{g,t-1}=0$, $\bm{D}^{-1}_{g,t}=\bm{D}^{-1}_{g,t-1}$ but there is no $g'$ such that $\bm{D}_{g',t} = \bm{D}_{g',t-1}=\bm{D}_{g,t-1}$, meaning that no other group experiences no treatment change and has the same baseline treatments as $g$. To identify the effect of the first treatment in one such cell, one could compare $g$'s outcome evolution to that of another group experiencing no change of its treatments and with different baseline treatments, as in the example in Equation \eqref{eq:example3}. However, such a comparison is only valid if the effects of the other treatments are constant over time, as discussed in Section \ref{subsub:intuition}.

\section{Alternative estimators of dynamic effects}
\label{web:dyn}

Assuming constant treatment effects, one can use a TWFE regression of the outcome on the treatment and its lags, the so-called distributed lag regression, to separately estimate the effect of the current and past treatments on the outcome. Separately estimating each of those effects while allowing for heterogeneous treatment effects is inherently difficult. This may be the reason why a substantial branch of the heterogeneity-robust DID literature has instead proposed to estimate the total effect of current and past treatments on the outcome \citep[see][]{callaway2018,abraham2018,de2020difference,borusyak2021revisiting}. We first show in Subsection \ref{sub:single_treatment} that in the case of a single treatment, $\DIDM^f$ and $\DIDM^b$ can actually be used to estimate the effect of the contemporaneous treatment and of one treatment lag, controlling for other treatment lags, in the spirit of the distributed-lag regression. Then, focusing on the total effect of current and past treatments as in the aforementioned papers, we propose estimators when there are several binary and staggered treatments that have dynamic effects. 

\subsection{Single treatment} 
\label{sub:single_treatment}

With a single treatment $D^s_{g,t}$, $\DIDM^f$ can be used to estimate the effect of the current value of $D^s_{g,t}$, allowing for dynamic effects. Assume that $(D^1_{g,t},...,D^K_{g,t})=(D^s_{g,t},...,D^s_{g,t-(K-1)})$. Then, our potential outcome notation allows the current treatment and its first $K-1$ lags to affect the outcome, so $\DIDM^f$ is an estimator of the effect of the current value of $D^s_{g,t}$ robust to dynamic effects up to $K-1$ lags. This is an improvement over the $\DIDM$ estimator in \cite{dechaisemartin2020two}, which is not robust to dynamic effects, except with a binary and staggered treatment. To achieve some robustness to dynamic effects, $\DIDM^f$ restricts the estimation to groups that did not experience a treatment change from $t-K$ to $t-1$. For instance, with $K=2$ and $(D^1_{g,t},D^2_{g,t})=(D^s_{g,t},D^s_{g,t-1})$, the $\DIDM^f$ estimator compares groups with $(D^s_{g,t-2},D^s_{g,t-1},D^s_{g,t})=(0,0,1)$ to groups with $(D^s_{g,t-2},D^s_{g,t-1},D^s_{g,t})=(0,0,0)$, and groups with $(D^s_{g,t-2},D^s_{g,t-1},D^s_{g,t})=(1,1,0)$ to groups with $(D^s_{g,t-2},D^s_{g,t-1},D^s_{g,t})=(1,1,1)$. On the other hand, the $\DIDM^f$ estimator may not be used to estimate the effect of past treatments on the outcome. For instance, with $K=2$ and $(D^1_{g,t},D^2_{g,t})=(D^s_{g,t-1},D^s_{g,t})$, $\mathcal{S}_1$ is empty: for any group $g$ such that $(D^s_{g,t-2}\ne D^s_{g,t-1}=D^s_{g,t})$, there cannot exist another group $g'$ such that $(D^s_{g',t-2}=D^s_{g',t-1}=D^s_{g',t})$, and $(D^s_{g',t-2},D^s_{g',t-1})=(D^s_{g,t-2},D^s_{g,t-1})$. 

\medskip
The opposite applies to $\DIDM^b$: it may not be used to estimate the effect of the current treatment allowing for dynamic effects, but it may be used to estimate the effect of past treatments on the outcome. For instance, with $K=2$ and $(D^1_{g,t},D^2_{g,t})=(D^s_{g,t-1},D^s_{g,t})$, $\mathcal{S}_2$ is not empty: it contains all $(g,t)$ cells such that $D^s_{g,t-1}\ne D^s_{g,t}=D^s_{g,t+1}$, for which there exists another group $g'$ such that $D^s_{g',t-1}= D^s_{g',t}=D^s_{g',t+1}=D^s_{g,t+1}$. Then, $\DIDM^b$ is a weighted average of two types of DIDs. DIDs of the first type
compare the $t+1$ to $t$ outcome evolution between groups such that $D^s_{g,t-1}=1, D^s_{g,t}=0,D^s_{g,t+1}=0$ and groups such that $D^s_{g,t-1}=0, D^s_{g,t}=0,D^s_{g,t+1}=0$. DIDs of the second type compare the $t+1$ to $t$ outcome evolution between groups such that $D^s_{g,t-1}=1, D^s_{g,t}=1,D^s_{g,t+1}=1$ and groups such that $D^s_{g,t-1}=0, D^s_{g,t}=1,D^s_{g,t+1}=1$. If the current outcome only depends on the current treatment and its first lag, and if Assumption \ref{hyp:CT_alt2} holds for $(D^1_{g,t},D^2_{g,t})=(D^s_{g,t-1},D^s_{g,t})$, then $\DIDM^b$ is unbiased for the average effect of switching the treatment's first lag from $0$ to $1$ holding the current treatment fixed, across all $(g,t)$s in $\mathcal{S}_2$.

\medskip
Of course, assuming that the current outcome only depends on the current treatment and its first lag is restrictive. One could instead assume, say, that the current outcome only depends on the current treatment and its first two lags. Then, with $K=3$ and $(D^1_{g,t},D^2_{g,t},D^3_{g,t})=(D^s_{g,t-2},D^s_{g,t-1},D^s_{g,t})$, $\DIDM^b$ is unbiased for the average effect of switching the treatment's second lag from $0$ to $1$, holding the current treatment and its first lag fixed, across all $(g,t)$ cells in $\mathcal{S}_2$. $\mathcal{S}_2$ now becomes the set of all $(g,t)$ cells such that $D^s_{g,t-2}\ne D^s_{g,t-1}=D^s_{g,t}=D^s_{g,t+1}$ and for which there exists another group $g'$ such that $D^s_{g',t-2}=D^s_{g',t-1}=D^s_{g',t}=D^s_{g',t+1}=D^s_{g,t+1}$. $\mathcal{S}_2$ contains fewer cells with $K=3$ and $(D^1_{g,t},D^2_{g,t},D^3_{g,t})=(D^s_{g,t-2},D^s_{g,t-1},D^s_{g,t})$ than with $K=2$ and $(D^1_{g,t},D^2_{g,t})=(D^s_{g,t-1},D^s_{g,t})$: allowing more treatment lags to affect the outcome may be more plausible, but it may also result in less precise estimators, that apply to a smaller population. Note also that with dynamic effects up to $K-1$ treatment lags, $\DIDM^b$ can be used to estimate the effect of the $K-1$th lag, but it cannot be used to estimate the effect of earlier lags.

\medskip
Overall, $\DIDM^f$ and $\DIDM^b$ can be used for some but not for all purposes in the presence of a single treatment with dynamic effects.


\subsection{Multiple treatments} 
\label{sub:multiple_treatment}

In the paper, we implicitly assume that the treatments under consideration do not have dynamic effects, since the outcome of a unit at period $t$ only depends on her period-$t$ treatment, not on her previous treatments.\footnote{On the other hand and as discussed above, Theorems \ref{thm:main} and \ref{thm:extension} do apply to dynamic effect cases, when the other treatment variables in the regression are lags of the treatment.} When treatments can have dynamic effects, estimating the effect of a treatment controlling for other treatments is difficult. We propose an estimation strategy when there are two binary treatments, which both follow a staggered adoption design. For any $g\in\{1,...,G\}$, let
$F^1_{g}=\min\{t:D^1_{g,t}=1\}$
denote the first date at which group $g$ receives the first treatment, with the convention that $F^1_{g}=T+1$ if group $g$ never receives that treatment. Similarly, let
$F^2_{g}=\min\{t:D^2_{g,t}=1\}$
denote the first date at which group $g$ receives the second treatment, with the convention that $F^2_{g}=T+1$ if group $g$ never receives that treatment.

\begin{hyp}\label{hyp:stagg_designs_twotreatments}
	(Staggered design with two binary treatments) For all $(g,t)\in \{1,...,G\}\times\{2,...,T\}$, $D^1_{g,t}\in \{0,1\}$, $D^2_{g,t}\in \{0,1\}$, $D^1_{g,t-1}\leq D^1_{g,t}$, $D^2_{g,t-1}\leq D^2_{g,t}$, and $F^2_g\geq F^1_g$.
\end{hyp}

Assumption \ref{hyp:stagg_designs_twotreatments} requires that both treatments weakly increase over time, which means that once a group has switched from untreated to treated, it cannot switch back to being untreated. Assumption \ref{hyp:stagg_designs_twotreatments} also requires that groups start receiving the second treatment after the first. This is typically satisfied when the second treatment is a reinforcement of the first. Our running example will be that of a researcher seeking to separately estimate the effects of medical and recreational marijuana laws in the US: so far, states have passed the former before the latter, and none of the medical and recreational laws passed since the late 1990s have been reverted. Another example where Assumption \ref{hyp:stagg_designs_twotreatments} holds include voter ID laws in the US, where non-strict laws are typically passed before strict ones \citep[see][]{cantoni2019strict}. Another example are anti-deforestation policies, where plots of lands are typically put into a concession, and then some concessions get certified \citep[see][]{panlasigui2018impacts}.

\medskip
To allow for dynamic effects, we need to modify our potential outcome notation. For all $(\bm{d}^1,\bm{d}^2)\in \{0,1\}^{2T}$, let $Y_{g,t}(\bm{d}^1;\bm{d}^2)$ denote the potential outcome of group $g$ at period $t$, if her two treatments from period $1$ to $T$ are equal to $\bm{d}^1,\bm{d}^2$. This dynamic potential outcome framework is similar to that in \cite{robins1986new}. It allows for the possibility that groups' outcome at time $t$ be affected by their past and future treatments.

\medskip
Our estimator relies on the following assumptions.
\begin{hyp}
	(No Anticipation) For all $g$, for all $(\bm{d}^1,\bm{d}^2)\in \{0,1\}^{2T}$, $$Y_{g,t}(\bm{d}^1;\bm{d}^2)=Y_{g,t}(d^1_1,...,d^1_t;d^2_1,...,d^2_t).$$\label{hyp:no_antic}
\end{hyp}

\vspace{-0.5cm}
Assumption \ref{hyp:no_antic} requires that a group's current outcome do not depend on her future treatments, the so-called no-anticipation hypothesis. \cite{abbring2003nonparametric} have discussed that assumption in the context of duration models, and \cite{malani2015}, \cite{botosaru2018difference}, and \cite{abraham2018} have discussed it in the context of DID models.

\medskip
For any $j\in \{1,...,T\}$, let $\0_j$ and $\1_j$ denote vectors of $j$ zeros and ones, respectively. We also adopt the convention that $\0_0$ and $\1_0$ denote empty vectors. Hereafter, we refer to $Y_{g,t}(\0_T;\0_T)$ as group $g$'s -treated potential outcome at period $t$, her outcome if she never receives either of the two treatments. Our estimators rely on the following assumption on $Y_{g,t}(\0_T;\0_T)$.

\begin{hyp}\label{hyp:strong_exogeneity_dyn}
	(Independent groups, strong exogeneity, and common trends for the never-treated outcome) For all $t\geq 2$ and $g\in\{1,...,G\}$,
$E(Y_{g,t}(\0_T;\0_T)-Y_{g,t-1}(\0_T;\0_T)|\bm{D})$ does not vary across $g$.
\end{hyp}
Assumption \ref{hyp:strong_exogeneity_dyn} is an adaptation of Assumptions \ref{hyp:independent_groups}-\ref{hyp:common_trends} to the set-up we consider in this section, and where we allow for dynamic effects. 

\medskip
Under Assumption \ref{hyp:stagg_designs_twotreatments}, the estimators of instantaneous and dynamic treatment effects proposed in \cite{de2020difference} can still be used with two treatments, redefining the treatment as $\tilde{D}_{g,t}=D^1_{g,t}+D^2_{g,t}$. However, those estimators will average together effects of the first and of the second treatment. Estimating separately the effect of the first treatment is straightforward: one can just compute the estimators in \cite{callaway2018} or \cite{de2020difference}, restricting the sample to all $(g,t)$s such that $D^2_{g,t}=0$. In the marijuana laws example, to estimate the effect of medical marijuana laws, one can just restrict the sample to all state$\times$year $(g,t)$ such that state $g$ has not passed a recreational law yet in year $t$. The horizon until which dynamic effects can be estimated will just be truncated by the second treatment.

\medskip
Estimating separately the effect of the second treatment is more challenging but can still be achieved, under the following, supplementary assumption.

\begin{hyp}\label{hyp:hom_dyn_effects}
	(Restriction on the effect of the first treatment) For all $g\in\{1,...,G\}$, $j\in \{1,...,T\}$, and $t>j$, there exists $\lambda_{j,g}(\bm{D})$ and $\mu_{j,t}(\bm{D})$ such that
	$$E(Y_{g,t}((\0_{j-1},\1_{T-(j-1)});\0_T)-Y_{g,t}(\0_T;\0_T)|\bm{D})=\lambda_{j,g}(\bm{D})+\mu_{j,t}(\bm{D}).$$
\end{hyp}
Assumption \ref{hyp:hom_dyn_effects} requires that the effect of the first treatment evolves over time in the same way in every group: for any $t>j+1$,
$$E(Y_{g,t}((\0_{j-1},\1_{T-(j-1)});\0_T)-Y_{g,t}(\0_T;\0_T)|\bm{D})-E(Y_{g,t-1}((\0_{j-1},\1_{T-(j-1)});\0_T)-Y_{g,t-1}(\0_T;\0_T)|\bm{D}),$$
the difference between group $g$'s effect of being treated for $t-(j-1)$ and $t-1-(j-1)$ periods, should be the same in every group. Still, Assumption \ref{hyp:hom_dyn_effects} allows such treatment effects to vary in an unrestricted way with the number of time periods over which a group has been treated, and to vary with the time period at which the treatment was adopted. It also allows, to some extent, the treatment effect to vary across groups: groups' treatment effects can be arbitrarily heterogeneous at the first period where they start receiving the treatment, but the period to period evolution of that effect should then be the same in every group.

\medskip
To understand why that assumption is needed, let us go back to the marijuana law example. Without Assumption \ref{hyp:hom_dyn_effects}, a state passing a recreational marijuana law may start experiencing a different outcome trend than other states that have only passed a medical law, either because of the recreational law, or because  its evolution of the effect of the medical law differs from that in other states. In other words, that assumption is key to disentangle the effects of the two treatments, which is often of interest. Under the standard parallel trends assumption on the never-treated outcome (Assumption \ref{hyp:strong_exogeneity_dyn}), one could only estimate the combined effects of the two treatments.

\medskip
Though it is arguably strong, this assumption is partly testable, as we explain in more details below: it implies that groups that start receiving the treatment at the same time should then have the same outcome evolution until they adopt the second treatment. A violation of this assumption would lead our estimators to be upward (resp. downward biased) if the effect of the first treatment increases less (resp. more) in groups that adopt the second treatment than in groups that do not adopt it.

\medskip
\begin{hyp}\label{hyp:non_pathological_design}
(Non-pathological design) There exists $(g,g')\in\{1,...,G\}^2$ such that $F^1_{g}=F^1_{g'}$ and $1<F^2_{g}<F^2_{g'}$.
\end{hyp}
For any $f\in \{1,...,T\}$, let
$$\mathcal{G}_f=\{g\in \{1,...,G\}: F^1_{g}=f\}$$
denote the set of groups that started receiving the first treatment at date $f$.
Let
$$\mathcal{F}=\{f\in \{1,...,T\}: \exists (g,g')\in \mathcal{G}_f^2: 1<F^2_{g}<F^2_{g'}\}$$
be the set of dates such that at least two groups start receiving the first treatment at that date and start receiving the second treatment at different dates. Assumption \ref{hyp:non_pathological_design} ensures that $\mathcal{F}$ is not empty. For any $f \in \mathcal{F}$,
let
\begin{align*}
NT_f = \max_{g\in \mathcal{G}_f}F^2_{g} - 1
\end{align*}
be the last period at which at least one group that started receiving the first treatment at period $f$ has still not received the second treatment. Then, we let $$L_{nt,f} = NT_f-\min_{g\in \mathcal{G}_f:F^2_{g}\geq 2}F^2_{g}$$ denote the number of time periods between the first date at which a group that started receiving the first treatment at date $f$ starts receiving the second treatment, and the last date at which a group that started receiving the first treatment at date $f$ has not received the second treatment yet.
Note that $L_{nt,f}\geq 0$ for all $f\in \mathcal{F}$.
Let also $$L_{nt}=\max_{f \in \mathcal{F}} L_{nt,f}.$$
For any $\ell\in\{0,...,L_{nt}\}$, $f\in \mathcal{F}$ such that $NT_f \geq \ell+f+1$, and $t\in\{\ell+f+1,...,NT_f\}$,
let $$N^{f}_{t,\ell} = \sum_{g\in \mathcal{G}_f:F^2_{g}=t-\ell} N_{g,t}$$
denote the population at $t$ of groups that started receiving the first treatment at date $f$ and the second treatment $\ell$ periods ago at $t$, and such that at least one group also started receiving the first treatment at date $f$ and has not started receiving the second treatment yet at $t$.
Let
$$N_\ell=\sum_{f\in \mathcal{F}:NT_f\geq \ell+f+1} \; \sum_{t=\ell+f+1}^{NT_f}N^{f}_{t,\ell}$$
be the total population in groups reaching $\ell$ periods after they started receiving the second treatment at a date where there is still a group that started receiving the first treatment at the same date as their group and that has not received the second treatment yet. Across those groups, the average cumulative effect of having received the second treatment for $\ell+1$ periods while fixing the first treatment at its observed value is
$$\delta_{\ell} =  E\left[\sum_{f\in \mathcal{F}:NT_f\geq \ell+f+1} \sum_{t=\ell+f+1}^{NT_f} \sum_{g\in \mathcal{G}_f,F^2_{g}=t-\ell}\frac{N_{g,t}}{N_\ell}  Y_{g,t}(\bm{D}^1_g;(\0_{t-\ell-1},\1_{\ell+1}))-Y_{g,t}(\bm{D}^1_g;\0_t)\right].$$
Remark that by construction, $N_\ell>0$ for all $\ell\in\{0,...,L_{nt}\}$, so $\delta_{\ell}$ is well-defined for such $\ell$. Note also that $\delta_{\ell}$ does not include the effect of the second treatment for groups that start receiving the two treatments at the same time. For those groups, it is impossible to separately estimate the effects of the first and second treatments, using our DID estimation strategy at least.

\medskip
We now define an estimator of $\delta_{\ell}$. For any $f\in\mathcal{F}$ and $t$ such that $NT_f \geq t$, let
$$N^{nt, f}_t = \sum_{g\in \mathcal{G}_f:F^2_{g}>t} N_{g,t}.$$
Then, for any $\ell\in \{0,...,L_{nt}\}$, $f\in\mathcal{F}$ such that $NT_f \geq \ell+f+1$, and $t\in\{\ell+f+1,...,NT_f\}$, we define
$$\DID^f_{t,\ell} = \sum_{g\in \mathcal{G}_f:F^2_{g}= t-\ell} \frac{N_{g,t}}{N^{f}_{t,\ell}}(Y_{g,t} - Y_{g,t-\ell - 1}) -
\sum_{g\in \mathcal{G}_f:F^2_{g} >t} \frac{N_{g,t}}{N^{nt,f}_t}(Y_{g,t} - Y_{g,t-\ell - 1})$$
if $N^{f}_{t,\ell}>0$ and $N^{nt,f}_t>0$, and we let $\DID^f_{t,\ell}=0$ if $N^{f}_{t,\ell}=0$ or $N^{nt,f}_t=0$. Then, for all $\ell \in \{0,...,L_{nt}\}$, we let
$$\DID_{\ell} = \sum_{f\in \mathcal{F}:NT_f\geq \ell+f+1}\; \sum_{t=\ell +f+1}^{NT_f}\frac{N^{f}_{t,\ell}}{N_\ell}\DID^f_{t,\ell}.$$

\begin{thm}\label{thm:dyn}
Suppose that Assumptions \ref{hyp:supp_gt} and \ref{hyp:stagg_designs_twotreatments}-\ref{hyp:non_pathological_design} hold. Then, $E\left[\DID_{\ell}\right]=\delta_{\ell}$ for all 
 $\ell\in\{0,...,L_{nt}\}$.
\end{thm}
$\DID_{\ell}$ can be computed by the \st{did\_multiplegt} Stata command, restricting the sample to the $(g,t)$s such that $D^1_{g,t}=1$, and including $F^1_g$ in the \st{trends\_nonparam} option. The asymptotic normality of the $\DID_{\ell}$ estimators, when the number of groups goes to infinity, could be established under similar assumptions and using similar arguments as those used to show Theorem 4 in \cite{de2020difference}.

\medskip
Beyond the somewhat complicated notation above, the idea underlying $\DID_{\ell}$ is actually quite simple: it amounts to comparing the outcome evolution of groups that adopt/do not adopt the second treatment, and that adopted the first treatment at the same date. This ensures that the ``treatment'' and ``control'' groups involved in this comparison have been exposed to the first treatment for the same number of periods. Under Assumptions \ref{hyp:strong_exogeneity_dyn} and \ref{hyp:hom_dyn_effects}, this in turn ensures that their outcome evolution would have been the same if the ``treatment groups'' had not adopted the second treatment. The estimation procedure we propose can easily be extended to more than two treatments. For instance, if there was a third treatment following a staggered adoption design and always adopted after the second one, one could estimate its effect using the \st{did\_multiplegt} Stata command, restricting the sample to the $(g,t)$s such that $D^2_{g,t}=1$, and including the interaction of $F^1_g$ and $F^2_g$ in the \st{trends\_nonparam} option.

\medskip
Theorem \ref{thm:dyn} complements the pioneering work of \cite{callaway2018} and \cite{abraham2018}, who provide DID estimators of the effect of a single treatment following a staggered adoption design. To our knowledge, our paper is the first to consider the case with several treatments following consecutive staggered designs, which arises relatively often, as the examples given above show. Our main insight is to show that one can obtain unbiased estimators of the effect of the second treatment, under the restriction on the effect of the first treatment stated in Assumption \ref{hyp:hom_dyn_effects}, and provided one controls for the first treatment's adoption date.

\medskip
The assumptions underlying Theorem \ref{thm:dyn} are refutable. They imply that groups that start receiving the treatment at the same time should have the same outcome evolution until they adopt the second treatment, see Equation \eqref{eq:CT_mod} in the Appendix. This can be tested, using similar placebo estimators as in \cite{de2020difference}, the main difference being that one should compare groups with the same value of $F^1_g$. The placebo estimators one can use to test the assumptions underlying Theorem \ref{thm:dyn} can also be computed by the \st{did\_multiplegt} command, restricting the sample to the $(g,t)$s such that $D^1_{g,t}=1$, including $F^1_g$ in the \st{trends\_nonparam} option, and requesting the \st{placebo} option. One should still keep in mind that such pre-trends tests come with some caveats, as shown by \cite{roth2019pre}: they may be underpowered and could fail to detect violations of the assumptions, and they may lead to pre-testing issues. However, our placebo estimators can be used to conduct the sensitivity analysis proposed by \cite{manski2018right} or \cite{rambachan2019honest}.

\medskip
The estimation strategy proposed in Theorem \ref{thm:dyn} requires that there is at least one pair of groups that receive the first treatment at the same date, and such that the first group receives the second treatment strictly before the second group. When the number of groups is relatively low (e.g.: the 50 US states), there may not be any pair of groups receiving the first treatment at the same time period. Then, two alternative estimation strategies can be proposed. First, instead of Assumption \ref{hyp:hom_dyn_effects}, one may assume that the effect of the first treatment evolves linearly with the number of periods of exposure, with a slope that differs across groups:
$$E(Y_{g,t}((\0_{j-1},\1_{T-(j-1)});\0_T)-Y_{g,t}(\0_T;\0_T)|\bm{D})=\lambda_{j,g}(\bm{D})+\mu_{g}(\bm{D})(t-j).$$
Then, one can recover the counterfactual outcome that a group adopting the second treatment would have obtained without it by extrapolating a linear estimate of its outcome evolution prior to adoption.
The resulting estimator can be computed by the \st{did\_multiplegt} command, restricting the sample to the $(g,t)$s such that $D^1_{g,t}=1$, and including the group indicator in the \st{trends\_lin} option.
Second, one could also strengthen Assumption \ref{hyp:hom_dyn_effects}, by assuming that the effect of the first treatment evolves potentially non-linearly with the number of periods of exposure, but that this evolution is the same in every group and at every time period:
$$E(Y_{g,t}((\0_{j-1},\1_{T-(j-1)});\0_T)-Y_{g,t}(\0_T;\0_T)|\bm{D})=\lambda_{j,g}(\bm{D})+\mu_{t-j}(\bm{D}).$$
Then, one can recover the counterfactual outcome that a group adopting the second treatment would have obtained without it, by extrapolating the outcome evolution experienced by groups that reached a similar number of periods of exposure to the first treatment without adopting the second one. The resulting estimator can be computed by the \st{did\_multiplegt} command, restricting the sample to the $(g,t)$s such that $D^1_{g,t}=1$, and including indicators for reaching $1$, $2$, etc. periods of exposure to the first treatment in the \st{controls} option.

\medskip
The approach in Theorem \ref{thm:dyn} can easily be extended to some instances where the assumptions of Theorem \ref{thm:dyn} fail. For example, there may applications with two binary treatments following a staggered adoption design, but such that some groups receive treatment $1$ before treatment $2$, other groups receive treatment $2$ first, and other groups receive both treatments at the same time. Then, one can start by restricting attention to the subsample of groups such that $D^1_{g,t}\geq D^2_{g,t}$ for all $t$ and $F^1_g<F^2_g$. This subsample includes groups that only receive treatment $1$, groups that receive both treatments but receive the second one strictly after the first, and groups that do not receive any treatment. In that subsample, one can estimate the instantaneous and dynamic effects of receiving only the first treatment, using the \st{did\_multiplegt} command and restricting the sample to $(g,t)$s such that $D^2_{g,t}=0$. One can then estimate the effect of receiving the second treatment when one has already received the first one, using the \st{did\_multiplegt} command, restricting the sample to the $(g,t)$s such that $D^1_{g,t}=1$ and including $F^1_g$ in the \st{trends\_nonparam} option. Second, one can restrict attention to the subsample of groups such that $D^2_{g,t}\geq D^1_{g,t}$ for all $t$ and $F^2_g<F^1_g$. In that subsample, one can estimate the effect of receiving only the second treatment, and the effect of receiving the first treatment when one has already received the second one, using the same steps as above but reverting the roles of the first of second treatments. Finally, one can restrict attention to the subsample of groups that either receive both treatments at the same time or that do not receive any treatment, and estimate the effect of receiving both treatments at the same time using the \st{did\_multiplegt} command. Comparing these five sets of estimates may be indicative of whether the treatments are complements or substitutes, even though differences could also be driven by heterogeneous effects across the various subsamples.

\section{Inference} 
\label{web:inf}

\subsection{Asymptotic approach} 
\label{sub:asymptotic_approach}

In this section, we show the asymptotic normality of $\DIDM^f$ and construct asymptotically valid (in fact, conservative) confidence intervals for $\delta_1$. The approach followed here is very similar to that considered in \cite{dechaisemartin2020two} and  \cite{de2020difference}. Also, note that the same reasoning directly applies to $\delta_2$ and $1/2[\delta_1+\delta_2]$.

\medskip
We use the same notation as in Section \ref{sub:inference} and the proof of Theorem \ref{thm:CI_ex}. Specifically, $s$ refers to an observed switch, associated with a final period $t_s$, an initial value of $D^1_{g,t}$, $d^1_s$, a final value of $D^1_{g,t}$, $d^1_s{}'$ and a value of $\bm{D}^{-1}_{g,t}$, $\bm{d}^{-1}_s$, constant between $t_s-1$ and $t_s$. Then, define
\begin{align*}
U_{G,g} & =  \frac{G}{\sum_{s=1}^S n_{1s} |d^1_s-d^1_s{}'|}\sum_{s=1}^S N_{g,t_s}
\sgn(d^1_s-d^1_s{}')\left[\indic{g\in \mathcal{G}_{1s}} -  \indic{g\in \mathcal{G}_{0s}} \frac{n_{1s}}{n_{0s} }\right] \Delta Y_{g,t_s},\\
\widehat{\sigma}^2 & = \frac{1}{G}\sum_{g=1}^G \left(U_{G,g} - \DIDM^f\right)^2.	
\end{align*}
The confidence interval of nominal level $1 - \alpha$  that we consider is
$$\CI^a = \left[\DIDM^f \pm z_{1-\alpha/2} \widehat{\sigma} G^{-1/2}\right],$$
where $z_{1-\alpha/2}$ is the quantile of order $1-\alpha/2$ of the standard normal distribution.

\medskip
Our results rely on Assumptions \ref{hyp:asym_design}-\ref{hyp:moment_cond} below. Hereafter, we let
$\Sigma_g:=V(Y_{g,1},...,Y_{g,T}|\bm{D})$ and let  $\underline{\rho}(\Sigma)$ denote the smallest eigenvalue of any symmetric semidefinite positive matrix $\Sigma$.

\begin{hyp}
\label{hyp:asym_design}
	(Asymptotically non-pathological design) The support of $\bm{D}_g$ does not depend on $g$ and is finite; it is denoted by $\mathcal{D}$. Let us define
	$$\mathcal{S}_g = \left\{(t, d^1, d^1{}', \bm{d}^{-1}): P(D^1_{g,t}=d^1,D^1_{g,t-1}=d^1{}', \bm{D}^{-1}_{g,t}=\bm{D}^{-1}_{g,t-1}=\bm{d}^{-1})>0 \right\}.$$
	Then $\mathcal{S}_g$ does not vary with $g$, $\mathcal{S}_g=\mathcal{S}$. Moreover, there exists $\underline{p}>0$ such that for all $g$ and $(t, d^1, d^1{}', \bm{d}^{-1})\in \mathcal{S}$, $P(D^1_{g,t}=d^1,D^1_{g,t-1}=d^1{}', \bm{D}^{-1}_{g,t}=\bm{D}^{-1}_{g,t-1}=\bm{d}^{-1}) \geq \underline{p}$ and $P(D^1_{g,t}=d^1{}',D^1_{g,t-1}=d^1{}', \bm{D}^{-1}_{g,t}=\bm{D}^{-1}_{g,t-1}=\bm{d}^{-1}) \geq \underline{p}$.
\end{hyp}

\begin{hyp}
\label{hyp:moment_cond}
(Regularity conditions for asymptotic normality) We have, for some $c>0$ and $\underline{\rho}>0$,
$$\sup_{g,t} N_{g,t}<\infty, \; \inf_{G, g\leq G} \underline{\rho}\left(\Sigma_g\right) \ge \underline{\rho} > 0 \;\; \text{a.s. and} \; \sup_{G, g\leq G, \bm{d}\in\mathcal{D}} E[|Y_{g,t}|^{2+c}|\bm{D}_g=\bm{d}]<\infty \;\; \text{a.s.}$$
\end{hyp}

Assumption \ref{hyp:asym_design} basically ensures that for all switch $s$ we observe, we have $n_{1s}\to\infty$ and $n_{0s}\to\infty$ almost surely as $G\to \infty$. Note that it always holds if groups are assumed to be identically distributed. Assumption \ref{hyp:moment_cond} ensures that we can apply the Lyapunov central limit theorem  with independent but not identically distributed variables. The second condition rules out degenerate situations where the (non-trivial) linear combinations of the $(Y_{g,1},...,Y_{g,T})$ in $\DIDM^f$ would actually be constant.

\begin{thm}\label{thm:asym}
	Suppose that Assumptions \ref{hyp:supp_gt}-\ref{hyp:independent_groups}, \ref{hyp:CT_alt} and \ref{hyp:asym_design}-\ref{hyp:moment_cond} hold. Then, conditional on $\bm{D}$ and almost surely,
$$	\sqrt{G}\frac{\DIDM^f - \delta_1}{\left(\frac{1}{G}\sum_{g:\bm{D}_{g,1}=0} V(U_{G,g}|\bm{D})\right)^{1/2}}  \convL \mathcal{N}(0,1),$$
Moreover, we have, almost surely,
	$$\liminf_{G\to\infty}  \Pr\left[\delta_1\in\CI^{\,a}|\bm{D}\right]\geq 1-\alpha.$$
\end{thm}


\subsection{Finite-sample inference}\label{sub:inference}

One limitation of the previous approach is that the asymptotic approximation may not be accurate. $\DIDM^f$ compares carefully selected treatment and control groups, and it could be the case that only a small number of groups can be included in those comparisons. The larger the number of treatments, the more likely it is that $\DIDM^f$ uses data from a small number of groups. In this section, we deal with this issue by proposing confidence intervals that are exact in a finite sample of groups under a normality assumption, in the spirit of \cite{donald2007inference}. The exactness of those confidence intervals relies on strong conditions,  but they remain asymptotically valid under much weaker assumptions. The main
price to pay for using them, rather than those of the previous section, is that doing so may result in an adjustment of the definition of $\delta_1$, as explained below.

\medskip
To ease the exposition, in this section we condition on $\bm{D}$. Accordingly, functions of $\bm{D}$ can be treated as non-stochastic terms. For simplicity, we also assume that $N_{g,t}=1$ for all $(g,t)$. Let $s=1,...,S$ index ``switches'', that is to say a $K+2$-uple $(d, d', \bm{d}^{-1},t)$ with $d \ne d'$ for which there exists $(g,g')$ satisfying $D^1_{g,t}=d$, $D^1_{g,t-1}=D^1_{g',t-1}=D^1_{g',t}=d'$ and $ \bm{D}^{-1}_{g,t}= \bm{D}^{-1}_{g,t-1}=\bm{D}^{-1}_{g',t}= \bm{D}^{-1}_{g',t-1}=\bm{d}^{-1}$. Then, note that
$$\DIDM=\sum_{s=1}^S \alpha_s \DID_s,$$
for some non-stochastic weights $(\alpha_s)_{s=1,...,S}$, where $\DID_s$ is the DID corresponding to switch $s$.  Let $\mathcal{G}_s$ denote the set of groups intervening in $\DID_s$, either as a ``switcher'' or as a ``control''.  We rely hereafter on the following assumption:
\begin{hyp}[Non-overlapping groups]\label{hyp:non_overlap}
	for any $(s,s')\in\{1,...,S\}^2$, $s\ne s'$, $\mathcal{G}_s\cap \mathcal{G}_{s'}=\emptyset$.
\end{hyp}
Combined with Assumption \ref{hyp:independent_groups}, this assumption ensures that the $\DID_s$ are mutually independent; otherwise it is unclear to us how one should account for their dependencies to still obtain a finite sample result. Note that Assumption \ref{hyp:non_overlap} automatically holds with $T=2$. Otherwise, it is more likely to hold if $T$ is small. When Assumption \ref{hyp:non_overlap} fails, we can ensure it holds on a modified sample, by removing groups belonging to several sets $\mathcal{G}_s$ from all those sets except one. This sample modification will modify the estimator $\DIDM^f$. It may also lead to removing a switching group from a set $\mathcal{G}_s$. This would change the target parameter, which would become the average treatment effect across all switching cells in the modified sample, in lieu of $\delta_1$, the average treatment effect across all switching cells in the original sample. For simplicity, we still denote the parameter and its estimator on the modified sample $\delta_1$ and  $\DIDM^f$.

\medskip
Our confidence interval relies on the following variance estimator:
$$\widehat{V} = \sum_{s=1}^S \alpha_s^2 \left[\frac{1}{n_{1s}(n_{1s}-1)}\sum_{g\in\mathcal{G}_{1s}} (\Delta Y_{g,t_s} -\overline{\Delta Y}_{1s})^2 + \frac{1}{n_{0s}(n_{0s}-1)} \sum_{g\in\mathcal{G}_{0s}} (\Delta Y_{g,t_s} -\overline{\Delta Y}_{0s})^2\right],$$
where $\mathcal{G}_{1s}$ (resp. $\mathcal{G}_{0s}$) is the subset of switching (resp. control) cells in $\mathcal{G}_s$, $n_{ks}=$card$(\mathcal{G}_{ks})$, and $\overline{\Delta Y}_{ks}$ is the average of $\Delta Y_{g,t_s}$ over $g\in\mathcal{G}_{ks}$. Our definition of $\widehat{V}$ uses the convention that 0/0=0.

\medskip
Next, let $q_{1-\alpha}$ denote the quantile of order $1-\alpha$ of $|T|$, defined as
\begin{equation}
T = \left(\frac{\sum_{s=1}^S \alpha^2_s \left(1/n_{1s}+1/n_{0s}\right)}{\sum_{s=1}^S \alpha_s^2 \left[W_{1s}/[n_{1s}(n_{1s}-1)]  + W_{0s}/[n_{0s}(n_{0s}-1)]\right]}\right)^{1/2} \times Z,	
	\label{eq:form_T}
\end{equation}
where $(Z, W_{01},W_{11},...,W_{0S},W_{1S})$ are independent of the data, mutually independent and satisfy $Z\sim\mathcal{N}(0,1)$ and $W_{ks}\sim \chi^2(n_{ks}-1)$. Note that $q_{1-\alpha}$ does not have a closed-form expression, but it can be approximated by simulations.

\medskip
Then, the confidence interval of order $1-\alpha$ we consider is
$$\CIex = \left[\DIDM\pm \, q_{1-\alpha} \sqrt{\widehat{V}}\right].$$
Below, we introduce two assumptions under which $\CIex$ is valid: under Assumption \ref{hyp:exact}, $\CIex$ is valid in finite samples; under Assumption \ref{hyp:asym}, $\CIex$ is valid asymptotically.
\begin{hyp}[Restrictions for finite-sample validity of $\CIex$]~\\[-0.5cm]
\begin{enumerate}
	\item For all $s=1,...,S$ and $g\in\mathcal{G}_s$, $Y_{g,t_s}(d_s^1,\bm{d}^{-1}_s) - Y_{g,t_s}(d_s^1{}',\bm{d}^{-1}_s) = \delta_{1s} $ where $(t_s,d_s^1,d_s^1{}',\bm{d}^{-1}_s)$ are the $(t, d^1, d^1{}', \bm{d}^{-1})$ associated with $s$ and $ \delta_{1s}$ is non-stochastic.
	\item For all $s$ and $g\in\mathcal{G}_s$, $\Delta Y_{g,t_s}(0,\bm{d}^{-1}_s) \sim \mathcal{N}(\mu_s, \sigma^2)$.
\end{enumerate}
\label{hyp:exact}
\end{hyp}
Point 1 of Assumption \ref{hyp:exact} assumes that the first treatment's effect is homogeneous within each set of groups $s$, but may vary across $s$. Point 2 ofAssumption \ref{hyp:exact} assumes that $\Delta Y_{g,t_s}(0,\bm{d}^{-1}_s)$ is normally distributed and homoskedastic: the variance of $\Delta Y_{g,t_s}(0,\bm{d}^{-1}_s)$ should not depend on $s$. When combined, the two points ensure that $\DID_s\sim\mathcal{N}(\delta_s, \sigma^2\alpha^2_s(1/n_{1s}+1/n_{0s}))$, which, in view of $\DIDM=\sum_{s=1}^S \alpha_s \DID_s$, is an important step towards our finite sample result below.

\begin{hyp}[Restrictions for asymptotic validity of $\CIex$]~\\[-0.5cm]
\begin{enumerate}
	\item There exists $G_0$ such that for all $G\ge G_0$, $\mathscr{S}_G:=\{(d,d',\bm{d}^{-1},t): N_{d,d',\bm{d}^{-1},t}>0,\, N_{d',d',\bm{d}^{-1},t}>0\}$ does not vary across $G$ and is finite. We denote by $\overline{S}$ its cardinal.\footnote{Accordingly, we keep the same indexation for switches $s\in\{1,...,\overline{S}\}$ for all $G\ge G_0$.}
	\item For all $(k,s)\in\{0,1\}\times\{1,...,\overline{S}\}$, the $(\Delta Y_{g,t_s})_{g\in\mathcal{G}_{ks}}$ are i.i.d. and for $g\in\mathcal{G}_{ks}$, $E[\Delta Y_{g,t_s}^2]<\infty$ and $V(\Delta Y_{g,t_s})>0$.
	\item For all $(k,s)\in\{0,1\}\times\{1,...,\overline{S}\}$, $\liminf_{G\to\infty} n_{ks}/G>0$.
\end{enumerate}
\label{hyp:asym}
\end{hyp}

The first point ensures that for $G$ large enough, $\DIDM$ can be written as a convex combination of $(\DID_s)_{s=1,...,\overline{S}}$, for the same type of switches $s$. Then, the last two points ensure that each of these $\DID_s$ is asymptotically normal and non-degenerate. Assumption \ref{hyp:asym} does not make any treatment-effect homogeneity or homoscedasticity assumption. Note that we impose that the $(\Delta Y_{g,t_s})_{g\in\mathcal{G}_{ks}}$ are identically distributed for simplicity. If we instead assumed that the $(\Delta Y_{g,t_s})_{g\in\mathcal{G}_{ks}}$ are independent but not identically distributed, one could still show that $\CIex$ is asymptotically conservative under appropriate regularity conditions, as in, e.g., Theorem S6 in the Web Appendix of \cite{dechaisemartin2020two}.

\medskip
The following theorem shows that $\CIex$ is exact under Assumption \ref{hyp:exact}, and asymptotically valid under Assumption \ref{hyp:asym}.
\begin{thm}\label{thm:CI_ex}
	If Assumptions \ref{hyp:supp_gt}-\ref{hyp:independent_groups}, \ref{hyp:CT_alt}, and \ref{hyp:non_overlap} hold, then, for any $\alpha\in (0,1)$:
\begin{enumerate}
	\item if  Assumption \ref{hyp:exact} further holds, $P\left(\delta_1\in \CIex\right)=1-\alpha$.
	\item if  Assumption \ref{hyp:asym} further holds, $\lim_{G\to\infty} P\left(\delta_1\in \CIex\right)=1-\alpha$.
\end{enumerate}
\end{thm}
Our approach in this section generalizes that in \cite{donald2007inference} to designs with more than two time periods and/or several treatments. A difference with that paper 
is that our confidence intervals use critical values from a non-standard distribution, instead of critical values from a t-distribution.


\section{Proof of the results in the Web Appendix}

\subsection{Theorem \ref{thm:dyn}} 
\label{sub:proof_of_theorem_ref_thm_dyn}

First, by Assumption \ref{hyp:strong_exogeneity_dyn}, for all $t\geq 2$ there is a function of $\bm{D}$ $\psi_t(\bm{D})$ such that
\begin{equation}
\psi_t(\bm{D})= E(Y_{g,t}(\0_T;\0_T) - Y_{g,t-1}(\0_T;\0_T)|\bm{D}).
\label{eq:CT_dyn}
\end{equation}
Then, for all $1\leq f<t\leq T$,
\begin{align}
& E[Y_{g,t}((\0_{f-1},\1_{T-f+1});\0_T) - Y_{g,t-1}((\0_{f-1},\1_{T-f+1});\0_T)|\bm{D}] \notag \\
= & E[Y_{g,t}((\0_{f-1},\1_{T-f+1});\0_T) - Y_{g,t}(\0_T;\0_T)|\bm{D}] - E[Y_{g,t-1}((\0_{f-1},\1_{T-f+1});\0_T) - Y_{g,t-1}(\0_T;\0_T)|\bm{D}]\notag \\
& + E[Y_{g,t}(\0_T;\0_T)- Y_{g,t-1}(\0_T;\0_T)|\bm{D}] \notag\\
= & \mu_{f,t}(\bm{D}) - \mu_{f,t-1}(\bm{D}) + \psi_t(\bm{D});
\label{eq:CT_mod}
\end{align}
where the second equality uses \eqref{eq:CT_dyn} and Assumption \ref{hyp:hom_dyn_effects}. Then, for any $\ell\in \{0,...,L_{nt}\}$, $f\in\mathcal{F}$ such that $NT_f \geq \ell+f+1$ and $t\in\{\ell+f+1,...,NT_f\}$ such that $N^{f}_{t,\ell}>0$ and $N^{nt,f}_t>0$,
\begin{align}\label{eq:thm_dyn_step1}
&E\left(\DID^f_{t,\ell}|\bm{D}\right)\nonumber\\
=& \sum_{g\in \mathcal{G}_f:F^2_{g}= t-\ell} \frac{N_{g,t}}{N^{f}_{t,\ell}}E\left(Y_{g,t} - Y_{g,t-\ell - 1}|\bm{D}\right)
 - \sum_{g\in \mathcal{G}_f:F^2_{g} >t} \frac{N_{g,t}}{N^{nt,f}_t} E\left(Y_{g,t} - Y_{g,t-\ell - 1}|\bm{D}\right)\nonumber\\
=&\sum_{g\in \mathcal{G}_f:F^2_{g}= t-\ell} \frac{N_{g,t}}{N^{f}_{t,\ell}}E\left(Y_{g,t}(D^1_g;(\0_{t-\ell-1},\1_{\ell+1})) - Y_{g,t}(D^1_g;\0_T)|\bm{D}\right)\nonumber\\
+&\sum_{g\in \mathcal{G}_f:F^2_{g}= t-\ell} \frac{N_{g,t}}{N^{f}_{t,\ell}}E\left(Y_{g,t}(D^1_g;\0_T) - Y_{g,t-\ell - 1}(D^1_g;\0_T)|\bm{D}\right)\nonumber\\
-&\sum_{g\in \mathcal{G}_f:F^2_{g} >t}\frac{N_{g,t}}{N^{nt,f}_t}E\left(Y_{g,t}(D^1_g;\0_T) -Y_{g,t-\ell - 1}(D^1_g;\0_T)|\bm{D}\right)\nonumber\\
=&\sum_{g\in \mathcal{G}_f:F^2_{g}= t-\ell} \frac{N_{g,t}}{N^{f}_{t,\ell}}E\left(Y_{g,t}(D^1_g;(\0_{t-\ell-1},\1_{\ell+1})) - Y_{g,t}(D^1_g;\0_T)|\bm{D}\right).
\end{align}
The first equality follows from the definition of $\DID^f_{t,\ell}$, and $N^{f}_{t,\ell}>0$ and $N^{nt,f}_t>0$. The second equality follows from Assumption \ref{hyp:no_antic}. The third equality follows from \eqref{eq:CT_mod}.

\medskip
By definition of $NT_f$, we have $N^{nt,f}_t>0$ for all $f\in\mathcal{F}$ and $t$ such that $NT_f \geq t$. We adopt the convention that a sum over an empty set is equal to 0. Then, for any $\ell\in \{0,...,L_{nt}\}$, $f\in\mathcal{F}$ such that $NT_f \geq \ell+f+1$ and $t\in\{\ell+f+1,...,NT_f\}$, Equation \eqref{eq:thm_dyn_step1} implies that
\begin{align*}
& N^{f}_{t,\ell} E\left(\DID^f_{t,\ell}|\bm{D}\right) \\
= & \sum_{g\in \mathcal{G}_f:F^2_{g}= t-\ell} N_{g,t} E\left(Y_{g,t}(D^1_g;(\0_{t-\ell-1},\1_{\ell+1})) - Y_{g,t}(D^1_g;\0_T)|\bm{D}\right).
\end{align*}
We obtain the result by summing over $f\in\mathcal{F}$ and $t$ such that $NT_f \geq \ell+f+1$ and $t\in\{\ell+f+1,...,NT_f\}$, and by the law of iterated expectations.

\subsection{Theorem \ref{thm:asym}} 
\label{sub:theorem_ref_thm_asym}

The proof is very similar to the proof of Theorem 3 in \cite{de2020difference} so we only discuss the main steps and stress the differences here.

\subsubsection*{1. Asymptotic normality} 
\label{ssub:asymptotic_normality}

First, remark that
$$\DIDM^f = \frac{1}{G}\sum_{g=1}^G U_{G,g}.$$
Then, we establish asymptotic normality by applying Lyapunov's central limit theorem for triangular arrays. To this end, we show
\begin{equation}
	\lim_{G\to\infty} \frac{\sum_{g=1}^G E\left[|U_{G,g} - E[U_{G,g}|\bm{D}] |^{2+c}| \; | \bm{D}\right]}{\left(\sum_{g=1}^G V(U_{G,g}\; | \bm{D})\right)^{1+c/2}}=0 \;\; \text{a.s.}
	\label{eq:cond_Lyap}
\end{equation}
We have $U_{G,g} = \frac{G}{\sum_{s=1}^S n_{1s} |d^1_s-d^1_s{}'|} \sum_{t=1}^{T} \lambda_{g,t} Y_{g,t}$ for some $\lambda_{g,t}$. Let
$$\underline{t}=\min\left\{t:\exists (d_1,d_1{}',\bm{d}^{-1}): (t,d_1,d_1{}',\bm{d}^{-1})) \in\mathcal{S}\right\}.$$
Let $s$ be a switch such that $t_s=\underline{t}$ and $(d^1_s, d^1_s{}', \bm{d}^{-1}_s)\in\mathcal{S}$. Then, if $g\in\mathcal{G}_{1s}$, we have $\lambda_{g,\underline{t}-1} = \sgn(d^1_s{}' - d^1_s)$. Then, reasoning as for obtaining (25) in \cite{de2020difference}, we get
\begin{equation}
\left(\frac{\sum_{s'=1}^S n_{1s'} |d^1_{s'}-d^1_{s'}{}'|}{G}\right)^2 \sum_{g=1}^G V(U_{G,g}\; | \bm{D})
\ge  \underline{\rho} \times n_{1s}.
	\label{eq:ineq1}
\end{equation}
Moreover, with the same reasoning as in \cite{de2020difference}, we have
$$\left(\frac{\sum_{s=1}^S n_{1s} |d^1_s-d^1_s{}'|}{G}\right)^2
E\left[|U_{G,g} - E[U_{G,g}|\bm{D}] |^{2+c}\; | \bm{D} \right] \leq C_0 \left[\max_{t=1...T} |\lambda_{g,t}|\right]^{2+c} $$
for some constant $C_0>0$. Moreover,
\begin{equation}
\max_{t=1...T} |\lambda_{g,t}| \le C_1 \sup_{g,t} N_{g,t}\left[1+\max_{s=1,...,S} \frac{n_{1s}}{n_{0s}}\right].	
	\label{eq:ineq2}
\end{equation}
We then obtain \eqref{eq:cond_Lyap} from \eqref{eq:ineq1} and \eqref{eq:ineq2} exactly as in the proof of Theorem 3 in \cite{de2020difference}.


\subsubsection*{2. Asymptotic validity of $\CI^{\,a}$} 
\label{ssub:asymptotic_validity_of_ci_a___1_allpha}

Let us define
$$\overline{\sigma}^2_{G}:= \frac{1}{G}\sum_{g=1}^G E\left[\left(U_{G,g} - \delta_1\right)^2|\bm{D}\right].$$
The exact same reasoning as in \cite{de2020difference} leads to
\begin{align}
\widehat{\sigma}^2 - \overline{\sigma}^2_{G}& \convP 0, \label{eq:conv_sig} \\
\overline{\sigma}^2_{G} &\geq \frac{1}{G}\sum_{g:\bm{D}_{g,1}=0} V(U_{G,g}|\bm{D}).
	\label{eq:ineg_sig}
\end{align}
Moreover, from \eqref{eq:ineq1}, Assumptions \ref{hyp:asym_design}-\ref{hyp:moment_cond} and the weak law of large numbers, we obtain that with probability approaching one,
$$	\frac{1}{G}\sum_{g=1}^G V(U_{G,g,\ell}|\bm{D})  \geq C_2 \underline{\rho}, $$
for some $C_2>0$. This ensures that $\left|\overline{\sigma}_{G}/\widehat{\sigma} - 1\right|\convP 0$. Then, write
$$G^{1/2}\frac{\DIDM^f-\delta_{1}}{\widehat{\sigma}} =
\frac{\left(\frac{1}{G} \sum_{g=1}^G V(U_{G,g}|\bm{D})\right)^{1/2}}{\overline{\sigma}_{G}} \times \left[\frac{\overline{\sigma}_{G}}{\widehat{\sigma}} \times
G^{1/2}\frac{\DIDM^f-\delta_{1}}{\left(\frac{1}{G} \sum_{g=1}^G V(U_{G,g}|\bm{D})\right)^{1/2}}\right].$$
By what precedes, the first term is smaller than one and the second converges to a standard normal variable. The result follows, with the same reasoning as in \cite{de2020difference}.


\subsection{Theorem \ref{thm:CI_ex}} 
\label{sub:theorem_ref_thm_ci_ex}

As in Subsection \ref{sub:inference}, the proof is conditional on $\bm{D}$.

\medskip
1. First, under Assumption \ref{hyp:exact}, we have $\delta_1 = \sum_{s=1}^S \alpha_s \delta_{1s}$. Thus, using again Assumption \ref{hyp:exact} but also Assumption \ref{hyp:non_overlap},
\begin{equation}
\DIDM-\delta_1=\sum_{s=1}^S \alpha_s (\DID_s - \delta_{1s}) \sim \mathcal{N}\left(0, \sigma^2\sum_{s=1}^S \alpha_s^2\left(\frac{1}{n_{1s}}+\frac{1}{n_{0s}}\right)\right).	
	\label{eq:DIDM_nor}
\end{equation}
For the same reasons, we have
$$\frac{\widehat{V}}{\sigma^2} \sim \sum_{s=1}^S \alpha_s^2 \left[\frac{W_{1s}}{n_{1s}(n_{1s}-1)}  + \frac{W_{0s}}{n_{0s}(n_{0s}-1)}\right],$$	
where the $(W_{01},W_{11},...,W_{0S},W_{1S})$ are mutually independent, independent of $\DIDM$ and $W_{ks}\sim \chi^2(n_{ks}-1)$, by Cochran's theorem. By definition of $T$ (see \eqref{eq:form_T}), this implies that
$$\frac{\DIDM-\delta_1}{\sqrt{\widehat{V}}} \sim T.$$
The result follows.

\bigskip
2. Without loss of generality, we assume hereafter that $G\ge G_0$, so that $\mathscr{S}_G$ does not vary across $G$ and has cardinal $\overline{S}$.

\medskip
Consider the ratio $R := (\DIDM-\delta_1)/\widehat{V}^{1/2}$. We first show that as $G \to\infty$, $R\convL \mathcal{N}(0,1)$. For any $(k,s)\in\{0,1\}\times\{1,...,\overline{S}\}$, let $\mu_{ks}:=E[\Delta Y_{g,t_s}]$ and $\sigma^2_{ks}:=V(\Delta Y_{g,t_s})$. Given that $n_{ks}\to \infty$, we have, by the central limit theorem,
$$\frac{\overline{\Delta Y}_{ks} - \mu_{ks}}{\sqrt{\sigma^2_{ks}/n_{ks}}} \convL \mathcal{N}(0,1).$$
Remark that $\DIDM=\sum_{s=1}^{\overline{S}} \alpha_s\left(\overline{\Delta Y}_{1s} - \overline{\Delta Y}_{0s}\right)$ and $\delta_1=\sum_{s=1}^{\overline{S}} \alpha_s\left(\mu_{1s} - \mu_{0s}\right)$. Moreover, $(\overline{\Delta Y}_{01}$, $\overline{\Delta Y}_{11},...,\overline{\Delta Y}_{0\overline{S}}, \overline{\Delta Y}_{1\overline{S}})$ are mutually independent by Assumptions \ref{hyp:independent_groups} and \ref{hyp:non_overlap}. Then, by, e.g., Lemma C.5 of \cite{dhtuvan2022}, we obtain
\begin{equation}
\frac{\DIDM-\delta_1}{\sqrt{\sum_{s=1}^{\overline{S}} \alpha_s^2\left(\sigma^2_{1s}/n_{1s}+\sigma^2_{0s}/n_{0s}\right)}}  \convL \mathcal{N}\left(0, 1\right).
	\label{eq:num_DID}
\end{equation}
Moreover, by the law of large numbers,
$$\frac{1}{n_{1s}-1}\sum_{g\in\mathcal{G}_{1s}} (\Delta Y_{g,t_s} -\overline{\Delta Y}_{1s})^2 \convP \sigma^2_{ks}.$$
Then, $\alpha_s^2\le 1$ and $\min_{k,s} \liminf_G n_{ks}/G >0$ implies that
\begin{equation}
G\left[\widehat{V}   -  \sum_{s=1}^{\overline{S}} \alpha_s^2\left(\sigma^2_{1s}/n_{1s}+\sigma^2_{0s}/n_{0s}\right)\right] \convP 0.	
	\label{eq:conv_var1}
\end{equation}
Next,
$$G\sum_{s=1}^{\overline{S}} \alpha_s^2\left(\sigma^2_{1s}/n_{1s}+\sigma^2_{0s}/n_{0s}\right) \ge
\left(\min_{k,s} \sigma^2_{ks}\right)\sum_{s=1}^{\overline{S}} \alpha_s^2 \ge \frac{\min_{k,s} \sigma^2_{ks}}{\overline{S}},$$
where the latter holds by convexity of $x\mapsto x^2$ and $\sum_{s=1}^{\overline{S}} \alpha_s=1$. Hence, by Assumption \ref{hyp:asym}, we obtain
$$\liminf_G G\sum_{s=1}^{\overline{S}} \alpha_s^2\left(\sigma^2_{1s}/n_{1s}+\sigma^2_{0s}/n_{0s}\right)>0.$$
Combined with \eqref{eq:conv_var1}, this yields
\begin{equation}
\frac{\widehat{V}}{\sum_{s=1}^{\overline{S}} \alpha_s^2\left(\sigma^2_{1s}/n_{1s}+\sigma^2_{0s}/n_{0s}\right)} \convP 1.
	\label{eq:denom_DID}
\end{equation}
Taken together, \eqref{eq:num_DID} and \eqref{eq:denom_DID} imply that $R\convL \mathcal{N}(0,1)$.

\medskip
Next, we prove that $T\convL \mathcal{N}(0,1)$. First, for all $(k,s)$, $n_{ks}\to\infty$ by Assumption \ref{hyp:asym}. Thus, by the law of large numbers, $W_{ks}/(n_{ks}-1)\convP 1$. In turn, using $\liminf_G n_{ks}/G >0$, we obtain
$$G\left[\sum_{s=1}^{\overline{S}} \alpha_s^2\left( \frac{W_{1s}}{n_{1s}(n_{1s}-1)} + \frac{W_{0s}}{n_{0s}(n_{0s}-1)}\right) - \sum_{s=1}^{\overline{S}} \alpha_s^2\left( \frac{1}{n_{1s}} + \frac{1}{n_{0s}}\right) \right] \convP 0.$$
Moreover, since $G/n_{ks}\ge 1$ for all $k,s$, we have
$$G \sum_{s=1}^{\overline{S}} \alpha_s^2\left( \frac{1}{n_{1s}} + \frac{1}{n_{0s}}\right) \ge 2 \sum_{s=1}^{\overline{S}} \alpha_s^2 \ge 1/\overline{S}.$$
As a result, $\liminf_G G \sum_{s=1}^{\overline{S}} \alpha_s^2\left( 1/n_{1s} + 1/n_{0s}\right)>0$. Hence,
$$\frac{\sum_{s=1}^S \alpha_s^2 \left[W_{1s}/[n_{1s}(n_{1s}-1)]  + W_{0s}/[n_{0s}(n_{0s}-1)]\right]}{\sum_{s=1}^S \alpha^2_s \left(1/n_{1s}+1/n_{0s}\right)} \convP 1.$$
Thus, by definition of $T$, $T\convL \mathcal{N}(0,1)$.

\medskip
By continuity of the normal distribution, this implies that $q_{1-\alpha}\to \Phi^{-1}(1-\alpha/2)$. Now, note that
$$P\left(\delta_1\in\CIex\right) = F_{|R|}(q_{1-\alpha}),$$
where $F_{|R|}$ denotes the cumulative distribution function of $R$, which converges to $x\mapsto \max(0,2\Phi(x)-1)$ by what precedes. Moreover, by P\'olya's theorem, the convergence is uniform. The  result follows.



\end{document}